\documentclass[pre,aps,10pt,twocolumn,twoside,floats,preprintnumbers,floatfix,superscriptaddress,a4paper,showpacs,showkeys,longbibliography]%
{revtex4-2}
\usepackage{amsmath}
\usepackage{amsfonts}
\usepackage{amssymb}

\usepackage{hyperref}
\usepackage[normalem]{ulem} 

\usepackage{url}

\usepackage{times}
\usepackage{time}
\usepackage{graphicx}

\usepackage{color}

\usepackage{color}

\usepackage{ulem}

\newcommand{\beq}{\begin{equation}}
	\newcommand{\eeq}{\end{equation}}

\def\ie{\emph{i.e.}~}

\def\fig#1{\ref{Fig:#1}}
\def\Fig#1{Fig.~\fig{#1}}

\begin{document}
        \title{Unifying  atoms and colloids near the glass transition through bond-order topology}
	
	\author{Laura Stricker}
	\email{laura.stricker@mat.ethz.ch}
	\affiliation{Department of Materials, ETH Z\"{u}rich, 8093 Z\"{u}rich, Switzerland}
	\author{Peter M. Derlet}
	\email{peter.derlet@psi.ch}
	\affiliation{Laboratory for Theoretical and Computational Physics, Paul Scherrer Institut, 5232 Villigen PSI, Switzerland}
        \author{Ahmet Faik Demir\"{o}rs}
        \affiliation{Department of Physics, University of Fribourg, 1700 Fribourg, Switzerland}
	\author{Hanumantha Rao Vutukuri}
	\affiliation{Active Soft Matter and Bio-inspired Materials Lab, Faculty of Science and Technology, University of Twente, MESA+ Institute, 7500 AE Enschede, The Netherlands}
	\author{Jan Vermant}
	\affiliation{Department of Materials, ETH Z\"{u}rich, 8093 Z\"{u}rich, Switzerland}
	

\begin{abstract}
In this combined experimental and simulation study, we utilize bond-order topology to quantitatively match particle volume fraction in mechanically uniformly compressed colloidal suspensions with temperature in atomistic simulations. The obtained mapping temperature is above the dynamical glass transition temperature, indicating that the colloidal systems examined are structurally most like simulated undercooled liquids. Furthermore, the structural mapping procedure offers a unifying framework for quantifying relaxation in arrested colloidal systems.
\end{abstract}

\maketitle

Colloidal systems have been successfully used as a convenient microscopic model for atomistic systems \cite{poo04}, as demonstrated by studies on both crystalline \cite{aue01} and glassy systems \cite{wee00,pha02}. Compared to their atomic counterparts, micrometer-sized systems offer the advantage of direct real-space visualization through microscopy imaging.
However, despite the large body of work based on the paradigm of "colloids-as-big-atoms" for equilibrium phase transitions \cite{ter15_BOOK,dho03_BOOK}, the equivalence has not been established as clearly for more challenging conditions, such as far-from-equilibrium or time-dependent dynamics. In this respect, the long-time dynamics close to the glass transition provides an interesting test case, that has been studied both theoretically, with the mode coupling theory (MCT) \cite{ben84}, and experimentally \cite{pus86,pus87}, and continues to be employed \cite{smu17,ver48,lin89,asa58}. Still, quantitative mapping of colloidal systems onto atomistic ones remains limited due to the ambiguity in determining the glass transition volume fraction \cite{bra09}.

The glass transition, both in colloidal suspensions and atomistic glass formers, is characterized by large increases in relaxation time  \cite{nar71} and viscosity \cite{mau09} as the system dynamically arrests \cite{jua08,tur17}. Interestingly, the microscopic dynamics are fundamentally different, i.e. Brownian versus Newtonian, as well as the dissipative mechanisms, since the interaction of colloids with the suspending medium dampens thermal fluctuations. Nevertheless, a nominally similar structural relaxation emerges \cite{got09_BOOK,bou11}, due to the time scale separation of slow local density fluctuations and the fast microscopic dynamics \cite{hun12,gle98,sza04,wee02}. Due to crowding, caging mechanisms appear through cooperative rearranging of particles with their nearest neighbours \cite{wee02}. Hence, a variety of short-range ordered arrangements originate that minimize the local potential energy, the so-called 'locally favoured structures' (LFS) \cite{tar05}. The development of LFS when approaching the glass transition is however limited by the geometric constraints posed by an Euclidean space \cite{roy15a}. In 3D, full tiling with maximally symmetric polyhedra (icosahedra) is not achievable and other defected topological structures emerge, such as the Frank-Kasper motifs \cite{fra58,fra59}. The situation is quite different in 2D, where minimally frustrated triangles can be tiled in a space filling way \cite{nel02_BOOK}, leading to fundamental differences between 2D and 3D glasses \cite{fle15}.

The geometric frustration, which occurs in 3D, originates from a competition between locally and globally favourable energetic configurations and ultimately prevents crystallization. Fragments of topologically close-packed Frank-Kasper (FK) phases have been observed in simulations both in undercooled liquids \cite{ped10} and in soft-particle systems \cite{iac11}. A comprehensive topological framework of the glass transition based on the homotopy of the icosahedral point-symmetry group was developed by Nelson \cite{nel83a,nel83b}, allowing for the enumeration of structural motifs based on the algebra associated to the SU(2) group, and has recently been successfully applied to describe simulated model atomistic systems \cite{der20a}. Compared to other LFS descriptions, such a framework provides an intrinsic link between local, mid-range and global ordering \cite{sad80}. This topological description is also able to reconcile the kinetic and the structural views of the glass transition. Recent work has shown that relaxation processes in binary model glasses are intimately related to the evolution of such structural motifs, where even the localized string-like structural excitation seen in both glasses and under-cooled liquids \cite{don98} can be rationalized in terms of a re-arrangement of the disclination network describing the defected topological structure \cite{der20,der21}.

In the present work, we exploit this structural perspective to establish a mapping between binary colloidal mixtures, compressed with an electric field (a system controlled by volume fraction), to atomistic simulations of an undercooled liquid undergoing temperature quenching. The agreement in the populations of structural motifs is excellent, suggesting their independence on the details of the interaction potential. In particular, our findings reveal that the 'arrested' experimental systems align with simulated temperatures exceeding the glass transition point. Slower compressions correspond to lower simulated mapping temperatures. In essence, our study provides a clear and unambiguous method to quantify relaxation in colloidal systems.

\paragraph*{Experiments and simulations. ---}
Experiments are carried out using a binary colloidal particle suspension whose concentration is controlled through dielectrophoretic forces generated by an electric field gradient ("dielectrophoretic bottle" \cite{PhysRevLett.96.015703}). A sketch of the experimental setup (Suppl.~Figs.~1 and 2) and additional details can be found in the Suppl.~Secs.~A-D. We use binary mixtures of sterically stabilized polymethyl methacrylate (PMMA) colloidal particles with diameters of 3.0~$\mu$m and 3.6$~\mu$m and a 5\% polydispersity, giving a ratio $\alpha$ = 0.84$\pm0.02$, and a dielectric constant $\epsilon_P \approx 2.6$ \cite{mik61}. The small and large particles are covalently labeled with the fluorescent dye 7-nitrobenzo-2-oxa-1,3-diazol (NBD) and Nile blue oxazone (Nile Red) respectively, and sterically stabilized with poly(12-hydroxystearic acid) \cite{bosma}. An additional radii ratio $\alpha$ = 0.7$\pm0.06$ is achieved with particle diameters 2.1 $\mu$m and 3 $\mu$m with a polydispersity of 4\% and 5\% respectively, where the large particles are dyed with rhodamine isothiocyanate (RITC). The particles are dispersed in equal weight parts in a mixture of cyclohexyl bromide (72.8~wt\%) and cis-decalin (27.2~wt\%), saturated with tetrabutylammonium bromide (TBAB) ($\epsilon_P$ = 5.6), to guarantee near- refractive index and density matching \cite{yethiraj-nature,vutukuri2012colloidal}. The final suspensions are equilibrated in the stock vial for 8 days, followed by another 3 days in the electric bottle sample cell, before turning the electric field on (frequency $f$ = 1~MHz). Different compression protocols are achieved by varying the intensity of the electric field ($E_{rms}$ = 0.2, 0.25 or 0.3 V$/\mu$m). Hence, we gradually increase the particle concentration by compression at constant stress, which entails a variable rate, starting from typically $\sim$20 wt\%. The structural evolution is resolved using confocal microscopy (Suppl.~Fig.~3).

The  experiments on the colloidal system are compared to molecular dynamics simulations of a model binary atomistic system described by pairwise potentials, optimally parameterized for the experimental systems. As the physics of bond frustration is mostly determined by the near-range inter-particle repulsion~\cite{Chandler}, we employ a modified version of the Wahnstr\"{o}m Lennard-Jones (LJ) potential \cite{wah91} whose bonding energy depends explicitly only on particle size. Instantaneous configurations of both the under-cooled liquid and the amorphous solid regime are obtained through a linear temperature quench from a well equilibrated high temperature liquid at fixed zero-pressure. Three different quench rates, spanning up to three-orders of magnitude, are used: $\dot{T}_{n}=\dot{T}_{0}/10^{n}$, with $n = 1,2,3$. Details of the LJ parameterization and the simulations are given in Suppl.~Sec.~E.

The local structural motifs, classified into 'local bonding classes' enumerated by Nelson~\cite{nel83b}, are identified in both experimental and simulated systems using a modified Voronoi tessellation \cite{der21,GITHUB}, by labelling the local environments via the triples $(N_{4},N_{5},N_{6})$. Here $N_n$ is the number of bonds of bond-order $n$. Two nearest-neighbour particles are identified as having a bond of order $n$ if they have $n$ common neighbours. Thus (0,12,0) represents the defect-free icosahedron, the minimally frustrated structure associated with both a low local energy and volume~\cite{fra52,cha78}. Further details on the analysis are provided in Suppl.~Secs.~H-I.

\begin{figure}[h]
	\includegraphics[width=0.95\linewidth]{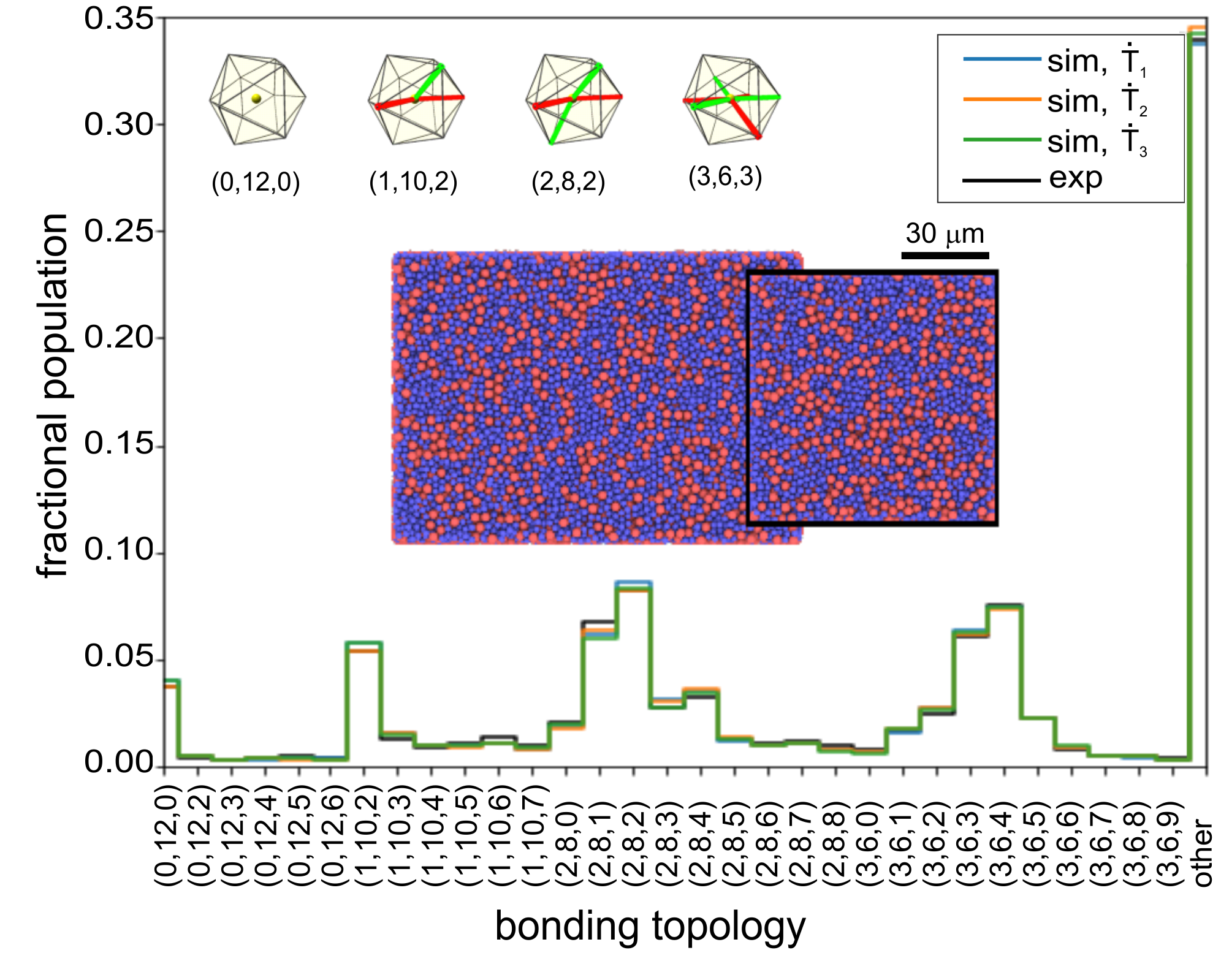}
	\caption{Comparison of the local structural motifs populations normalized by the total particle number, for a colloidal experiment and atomistic simulations of a thermally quenched LJ system at the effective temperatures $T_n^*$, for different quench rates $\dot{T}_{n}=\dot{T}_{0}/10^n$, with $n = 1,2,3$. The system has $\alpha = 0.7$ and it is experimentally compressed with $E_{rms} = 0.25$ V/$\mu$m up to a volume fraction $\Phi\sim0.52$. The inset depicts schematics of most frequently occurring bonding topologies, with green and red lines indicating bonds of order 4 and 6 respectively, and a planar view of experiments (left) and simulations at $\dot{T}_{3}$ (right).}
	\label{Fig:Nelson classes, sim-exp}
\end{figure}

\paragraph*{Local structural motifs and mapping temperature for athermal experiments. ---}
The local structural motifs in the experimental configurations achieved by compression of the colloidal systems are compared here to those in the simulated atomistic system, undergoing thermal quenching. In simulations, the glass transition regime manifests itself by a change in slope in the energy and volume per particle with respect to decreasing temperature (Suppl.~Fig.~5), where the intersection of the low and high temperature linear extrapolations gives a well-founded estimate of the glass transition temperature $T_{\mathrm{g}}$ (Suppl.~Fig.~7). At temperatures just above $T_{\mathrm{g}}$, all curves for different $\dot{T}_{n}$ progressively deviate from each other, indicating kinetic arrest and the transition to a structural glass. The slowest quench ($n=3$) produces the lowest glassy cohesive energy and volume per particle, and hence the most relaxed structure. Correspondingly, the icosahedral/Frank-Kasper (IFK) fraction content increases with decreasing temperatures, as the system enters the glass transition regime (Suppl.~Fig.~4). Slower quench rates (i.e more relaxed glasses) have an increased IFK fraction. This agrees with earlier works showing that well relaxed glassy structures consist of a kinetically arrested system-spanning network of small particle icosahedral motifs~\cite{jon88,she06,ma15} penetrated by 6-fold defect bonds associated with large particle Frank-Kasper structures~\cite{fra58}, with the remaining regions consisting of 4-fold, 5-fold and 6-fold bonds~\cite{der20}. On the other hand, in a deeply under-cooled liquid regime, before the glass transition, the structure generally consists of non-percolating icosahedral motives. Due to their topological origin, such low-energy structural motifs are rather insensitive to local distortions. Hence, they offer a platform to map the experimental colloidal systems onto the simulated atomistic ones, revealing how relaxed the colloidal systems really are.

\begin{figure}
	\includegraphics[width=1.1\linewidth]{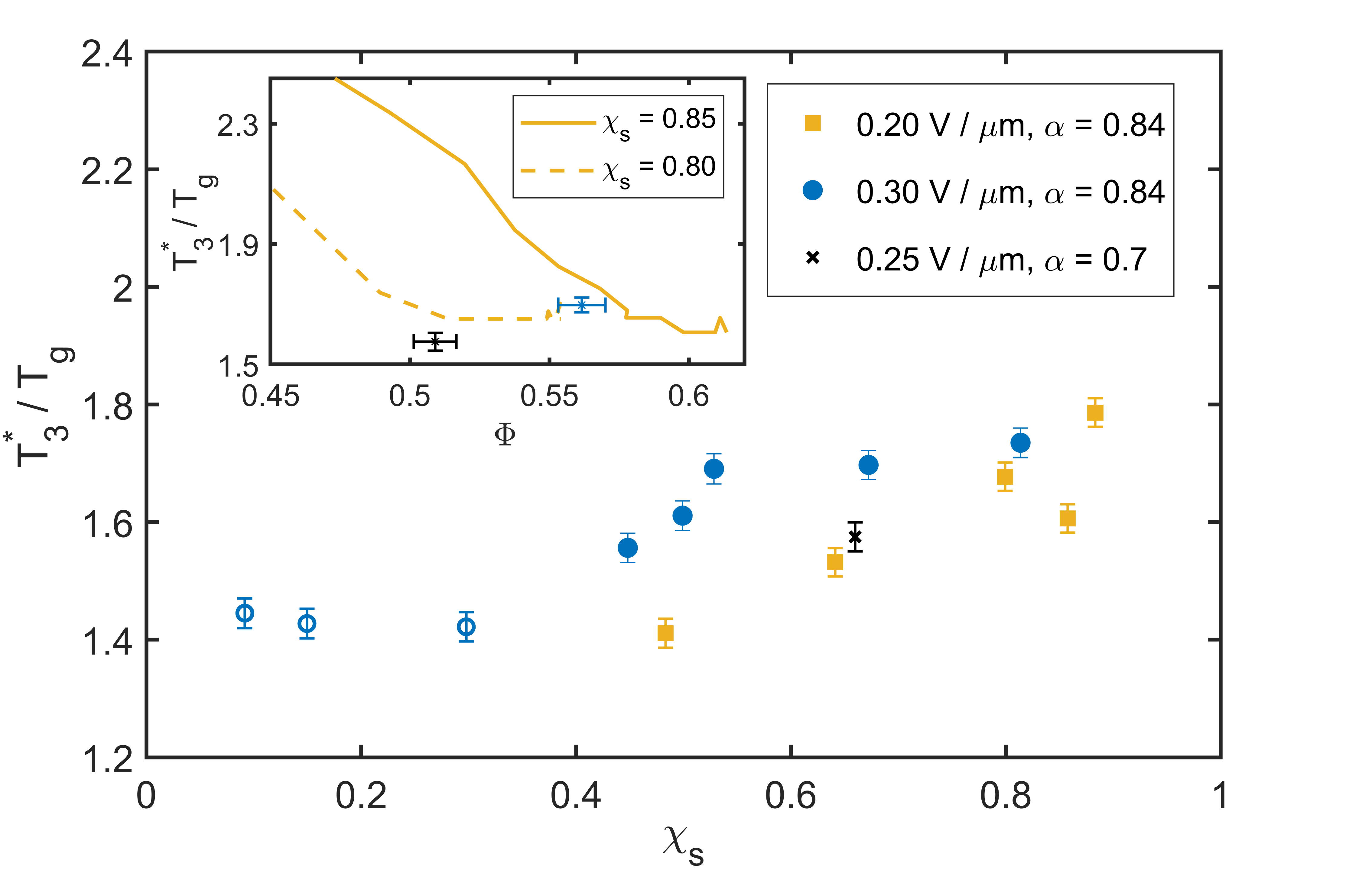}
	\caption{Normalized effective temperature of compressed colloidal systems versus composition. The systems are compressed with $E_{rms} = 0.2~V/\mu$m (yellow), $0.25~V/\mu$m (black) and $0.3~V/\mu$m (blue) for 5 days (filled symbols) or 52 hours (open symbols). The insert depicts the evolution of the effective temperature versus volume fraction, during a compression, for systems with $\chi_s=0.8$ (dashed line) and $\chi_s=0.85$ (solid line). The two data points (blue and black) correspond to two different realizations at $\chi_s=0.75$, for which we provide the detailed simulated dynamics (Suppl.~Figs.~13 and 14).}
    \label{Fig:Map exp-sim}
\end{figure}

To map an instantaneous configuration of the colloidal system onto the thermal atomistic one, for each quench rate $\dot{T}_n$ we determine an effective temperature $T_n^*$, based on the optimal quantitative matching of the experimental and simulated $(N_{4},N_{5},N_{6})$ populations, as defined by their minimum root-mean-square (RMS) residual (Suppl.~Fig.~6). The corresponding populations of local bonding motifs found in experiments and simulations at $T_n^*$ are displayed in \Fig{Nelson classes, sim-exp} and Suppl.~Fig.~11 for two different samples, revealing a remarkable agreement. The combination of the well founded estimate of $T_g$ from simulations with the mapping of the colloidal system on the atomistic one provides an internally consistent way of unifying colloidal and atomistic systems, although they are quenched in different ways.

\paragraph*{Effect of compression protocol and composition. ---}
We repeat the mapping procedure for consecutive configurations achieved during an experimental compression, to obtain a time-resolved mapping while approaching the glass transition for a system at a given composition $\chi_s = V_{s}/(V_{s} + V_{l})$, where $V_{s}$ and $V_{l}$ are the total volumes of small and large particles respectively. In the electric bottle experiment, suspensions undergo compression under a constant stress, possibly slightly reduced upon crowding. \Fig{Map exp-sim}'s insert displays the effective temperature as a function of the experimental volume fraction $\Phi$ for two continuous compression experiments, where time is an implicit variable. Over longer times, the curves level off and reach a plateau. This is attributed to a slowing down in structural rearrangements. Upon compression or upon quenching,  an increasing number of low-energy structural motifs emerge (Suppl.~Fig.~10), leading to increasing energy barriers for further structural evolution and prolonged relaxation times \cite{fra52}. Our athermal experimental systems can only explore such motifs when the barrier energies are small enough to be overcome by stress-driven structural instabilities during constant stress compression in the experiments. Once these low-energy structural transitions become exhausted, the evolution becomes exceedingly slow, seemingly arrested within the experimental time frame. However, the mapping method enables verification of whether the system has effectively entered a glassy state.
The volume fraction at which the plateau is reached in the experiments is approximately $\Phi_p\sim$~0.5 for $\chi_s$ = 0.8 and $\Phi_p\sim$~0.6 for $\chi_s$ = 0.85, with initial volume fraction of 0.25 and 0.3 respectively. Notably, even a minor composition variation of 6\% results in a more than 15\% variation in $\Phi_p$. This variation exceeds typical experimental uncertainties in volume fraction measurement, which are around 3\% \cite{poo12}. Therefore, we conclude that, for a given compression protocol, $\Phi_p$ also depends on the initial volume fraction, in line with previous findings \cite{lib13}, while the mapping temperature of the final slowed-down state primarily depends on the composition $\chi_s$.

In \Fig{Map exp-sim}, we display the normalized effective temperature $T_3^*/T_g$ as a function of composition $\chi_s$. These results were achieved by compressing for 5 days with field intensities $E_{rms}$ = 0.2 V/$\mu$m (yellow) and $E_{rms}$ = 0.3 V/$\mu$m (blue).
Normalizing $T_3^*$ with respect to $T_g$ offers a quantifiable measure of proximity to the glass transition and, consequently, the relaxation of the colloidal system. As expected, the weaker hence slower compression (yellow) leads to lower effective temperatures $T_3^*$/$T_g$, indicating a closer proximity to the glass transition. The mapping temperature $T_n^*$ weakly depends on the simulated quench rate, where faster quench rates correspond to lower $T_n^*$ values (Suppl.~Figs.~6 and 8). Such a weak dependence reflects the close-to meta-equilibrium state of the simulated undercooled liquid. By employing the mapping protocol, we can also incorporate results where the compression is halted after 52 hours (represented by open circles) into an effective measure of the distance to the glass transition. Remarkably, all mapping temperatures are well above the corresponding glass transition temperature $T_{\mathrm{g}}$, and thus in the meta-equilibrium of the simulated under-cooled liquid. This observation is also confirmed by the fact that the corresponding simulated systems have not yet developed a caging plateau in the intermediate scattering function ISF \cite{rei05} (Suppl.~Fig.~13), nor a divergence in the dynamic viscosity $\eta$ (Suppl.~Fig.~14). Details of the calculation of the ISF and $\eta$ are provided in Suppl. Sect. G. For both compression protocols, the mapping temperature increases with the fraction of small particles $\chi_s$. At increasing fractions of small particles several factors hinder the overcoming of energy barriers: a larger contact area, entailing higher dissipation in over-damped dynamics, higher volume filling of the small particles \ie higher crowding, and lower heterogeneity. Higher degrees of heterogeneity are indeed associated to more unstable energy states, with more fluctuating energy landscapes favouring relaxation \cite{wan19}. This is in line with previous numerical findings that localized stress-driven structural transitions (shear transformation zones \cite{fal11}) can mediate plasticity in quasi-static loading protocols due to local volume and therefore stress heterogeneity. Hence heterogeneity can play the role of an effective temperature in an athermal amorphous solid \cite{fal98}. To additionally assess the role of particle sizes, we consider a system with particle diameters 2.1 $\mu$m and 3 $\mu$m, namely a ratio $\alpha=0.7$, compressed with $E_{rms} = 0.25 \;V/\mu$m (black cross). Compared to the system with $\alpha=0.84$, the mapping temperature in the arrested state is lower, indicating a greater degree of relaxation. Once more, we attribute this observation to the increased heterogeneity of the system with $\alpha$ = 0.7.

\begin{figure}
	\includegraphics[width=1\linewidth,trim=0cm 0cm 0cm 0cm]{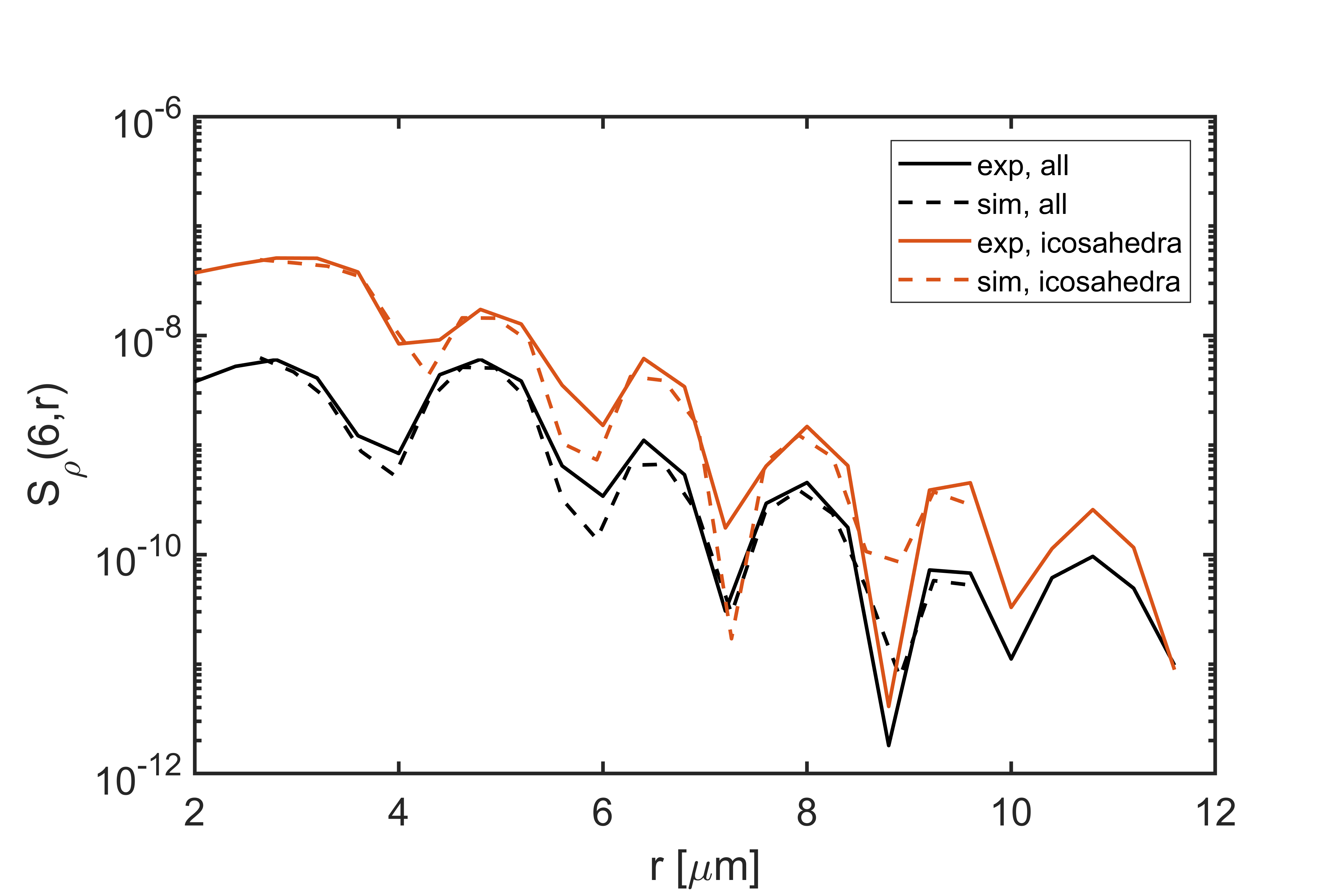}
	\caption{Square root of the angular power spectrum $l=6$ versus radial distance, for all particles (black) and for icosahedral-centered particles (red), in experiments (solid lines) and simulations (dashed lines), for the system of \Fig{Nelson classes, sim-exp}.}
	\label{Fig:NewFig3}
\end{figure}

\paragraph*{Mid-range icosahedral ordering. ---}
The robustness of the icosahedral motif has long been appreciated at the nearest-neighbours level and used to investigate the LFS structures of under-cooled liquids \cite{cel07,leo12,taf13,roy17,tur18,tan19,win19}. Icosahedral symmetry has however also been used to evaluate mid-range ordering. A recent probe  exploits the spherical harmonic transform of a 4-point spatial correlation function, where the $l=6$ harmonic coefficient is found to dominate, indicating a non-negligible icosahedral point-like symmetry of liquid medium-range order \cite{yua21,sin23}. \Fig{NewFig3} and Suppl.~Fig.~12 display a similar analysis for two  experiments and the corresponding simulations. The details of the analysis are described in Suppl. Sect.~I. In particular, we plot $S_{\rho}(l,r)=\left[(2l+1)^{-1}\sum_{m=-l}^{l}|\rho_{l}^{m}(r)|^{2}\right]^{1/2}$ for $l=6$ as a function of the radial distance $r$, with $\rho(r,\theta,\varphi)=\sum_{l=0}^{\infty}\sum_{m=-l}^{l}\rho_{l}^{m}(r)Y_{l}^{m}(\theta,\varphi)$ the particle density around a particle averaged over all such particles, decomposed with respect to the spherical harmonics $Y_{l}^{m}(\theta,\varphi)$, where $\theta,\varphi$ are the angular coordinates. We find a close agreement between the colloidal and model LJ system at the mapping temperature. If the particle at the origin has the icosahedral motif, the $S_{\rho}(6,r)$ increases by an order of magnitude, demonstrating the relevance of non-icosahedral motifs, but also the Frank-Kasper motifs which accommodate its defected packing. Such a structural indicator has been directly related to dynamic heterogeneities \cite{sin23}. Therefore, the agreement in mid-range ordering does not only confirm the validity of our mapping approach, but also suggests a direct link between the proposed local motif analysis and some dynamical aspects of the glass transition. It is important to underline that, whilst inertial and stochastic dynamics can give similar results in this temperature regime \cite{gle98,sza04,hij17}, the over-damped Brownian dynamics of our experimental system will give an entirely different time-scale for arrest. Therefore, our structural mapping does not directly translate into a dynamic mapping of colloidal and atomistic systems. Still, it allows to establish a meaningful baseline to study deviations from hard-sphere Brownian dynamics "at equal structure", which can be calculated as described in Suppl. Sect.~G. In colloidal systems such deviations are expected due to the presence of brush layer lubrication \cite{nom12}, hydrodynamics \cite{gra19}, hydrodynamics within the brush \cite{ohn10}, softness \cite{roy13} and non central or contact forces \cite{ngu20}.

\paragraph*{Conclusions. ---}
A colloidal glass-former, arrested by mechanically increasing the volume fraction and brought close to the glass transition, has been compared to a thermally quenched atomistic simulated LJ system. The striking similarity in the bonding topology distributions of local structural motifs provides a robust method for establishing a quantitative equivalence between particle volume fraction and temperature, eliminating experimental ambiguities. Specifically, this method demonstrates that our arrested colloidal systems exhibit structural similarities to deeply under-cooled liquids, even at high volume fractions $0.58\leq \Phi\leq0.64$, where $\Phi$ = 0.58 is typically considered the glass transition limit for near-hard colloids \cite{hun12} and $\Phi$ = 0.64 the maximum random packing fraction \cite{li10}. The consistency of our approach is evident in the excellent agreement between mapped thermal and athermal systems, particularly regarding mid-range ordering, assessed by deviations from icosahedral point-group symmetry. These deviations, stemming from the connectivity of local structural motifs \cite{nel02_BOOK}, are closely linked to dynamical heterogeneities \cite{sin23}. Overall, we provide a way to unambiguously determine how close to the glass transition a colloidal system is by quantitatively assessing its relaxation.

\section*{Acknowledgements}
We thank Prof. Andr\'{e} Studart's lab (ETH, Z\"{u}rich) for allowing us to access the confocal microscope and electric field setups, in particular Marco Binelli (ETH, Z\"{u}rich) for his support with the experimental setup. We are thankful to Johan Stiefelhagen (Utrecht University) and Andrew Schofield (Univeristy of Edinburgh), who kindly provided the colloidal particles. J. V. acknowledges the late W. R. Russel (Princeton University) for illuminating discussions. We acknowledge financial support from the Swiss National Science Foundation SNSF (project 192336).

\clearpage

%

\clearpage

\section*{Supplementary Material}
\renewcommand{\figurename}{Suppl. Fig.}
\setcounter{figure}{0}

\subsection{Colloidal suspensions}
Our experimental system consists of binary mixtures of poly(methyl methacrylate) (PMMA) particles covalently labeled with the fluorescent dye 7-nitrobenzo-2-oxa-1,3-diazol (NBD) or Nile blue oxazone (Nile Red) and sterically stabilized with poly(12-hydroxystearic acid) \cite{bosma}. The smaller particles have a diameter of $\sigma_{S} = 3.008 ~\mu$m, while the larger particles have a diameter of $\sigma_{L} = 3.576~ \mu$m, both with a polydispersity of $5\%$, entailing a diameter ratio $\alpha = 0.84\pm0.02$. Alternatively, a mixture with $\sigma_{S} = 2.1~\mu$m and $\sigma_{L} = 3~\mu$m with a polydispersity of $4\%$ and $5\%$ respectively ($\alpha = 0.7\pm0.06$) is used, where the large particles are dyed with rhodamine isothiocyanate (RITC) instead of Nile Red. The particles are suspended in a 20\% mixture of cyclohexyl bromide (72.8 wt\%) and cis-decalin (27.2 wt\%), saturated with tetrabutylammonium bromide (TBAB). In such a solvent mixture, the particles are nearly density and refractive index-matched, and they behave like near-hard spheres \cite{yethiraj-nature,anand-review}. We use mixtures with different compositions $\chi_s = V_s/(V_s + V_l)$, in the range $\chi_s = 0.1-0.9$, where $V_s$ and $V_l$ are the total volumes of the small and large particles respectively.

\subsection{Dielectrophoretic bottle}
	
Dielectrophoresis is the translational motion that polarizable objects exhibit as a result of their interaction with non-uniform electric fields. Particles whose dielectric constant $\epsilon_p$ is different from the one of the suspending medium, $\epsilon_m$, acquire a dipole moment. Hence, they are either attracted towards ($\epsilon_p > \epsilon_m$, ‘positive’ dielectrophoresis), or repelled
from ($\epsilon_p < \epsilon_m$, ‘negative’ dielectrophoresis) the areas with the highest electric field. The balance between the osmotic pressure and the electrostatic driving force ensures full equilibrium \cite{PhysRevLett.96.015703}.
When there is a field gradient, colloidal particles experience a dielectrophoretic force:
	
\begin{equation}\label{dep-force}
\mathbf{F}_{dep} = - \frac{1}{2}\nu_p \epsilon_{eff} \epsilon_{\circ} \nabla\mathbf{E}^{2}(r)
\end{equation}
	
\noindent where $\epsilon_\circ$ is the permittivity of vacuum, $\nu_p$ is the volume of the particle, $\bf{E}$ is the local field strength, and $\epsilon_{eff}$ is the effective dielectric constant of the particle in the suspension. The effective particle dielectric constant $\epsilon_{eff}$ is a function of the complex frequency-dependent dielectric constants of the particle and the solvent. Moreover, it also depends on the overall particle concentration, because the local field strength around a particle is altered by the induced dipole moments of the surrounding particles. The effective dielectric constant can be expressed by the Clausius-Mossotti equation \cite{PhysRevLett.96.015703}, $\epsilon_{eff} = 3 \alpha \epsilon_{m}/(1-\alpha \phi)^2$, where $\phi$ is the volume fraction of the particles and $\alpha$ is the real part of the Clausius-Mossotti factor. For the field frequencies used here, $\alpha$ is given by:
	
\begin{equation}\label{beta}
    \alpha =  \frac{(\epsilon_p-\epsilon_m)}{(\epsilon_p + 2 \epsilon_m)}.
\end{equation}
		
\begin{figure}
\includegraphics[width=0.9\linewidth]{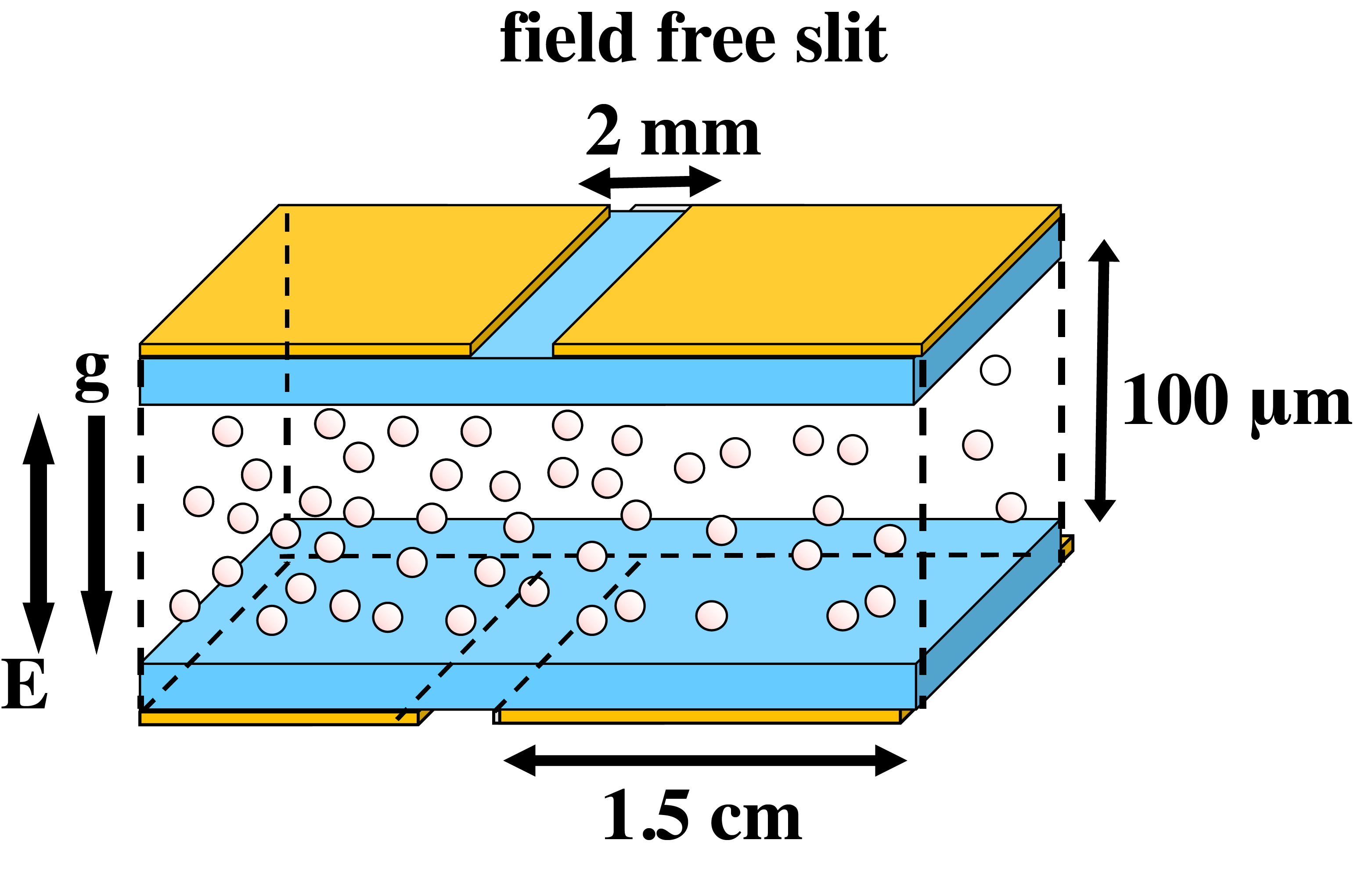}
\caption{Schematic representation of the e-bottle sample cell. The sputtered-coated electrodes are depicted in yellow. The respective directions of gravity
and the applied (AC) electric field when the cell was mounted on the stage of the microscope are also shown. We note that the dimensions in the cartoon are not scaled with actual size.}
\label{Fig:e-setup}
\end{figure}
	
In our setup, we choose to use a dielectrophoretic bottle to achieve a slow compression of the colloidal sample. An alternative way to experimentally study the aging behavior of a glass-forming system would be to quench the sample from a liquid state. This could be achieved in two ways: i) by rapidly lowering the temperature or by increasing the pressure, for atomic or molecular systems, and ii) by applying centrifugal forces for colloidal near-hard spheres. However, rapid quenching of the system can give rise to unwanted stress and strain fields in the sample. The usage of a slow compression protocol achieved by dielectrophoretic forces allows us to minimize the presence of remaining stress and strain fields.
	
\subsection{Experimental setup}
\begin{figure}
\includegraphics[width=0.9\linewidth]{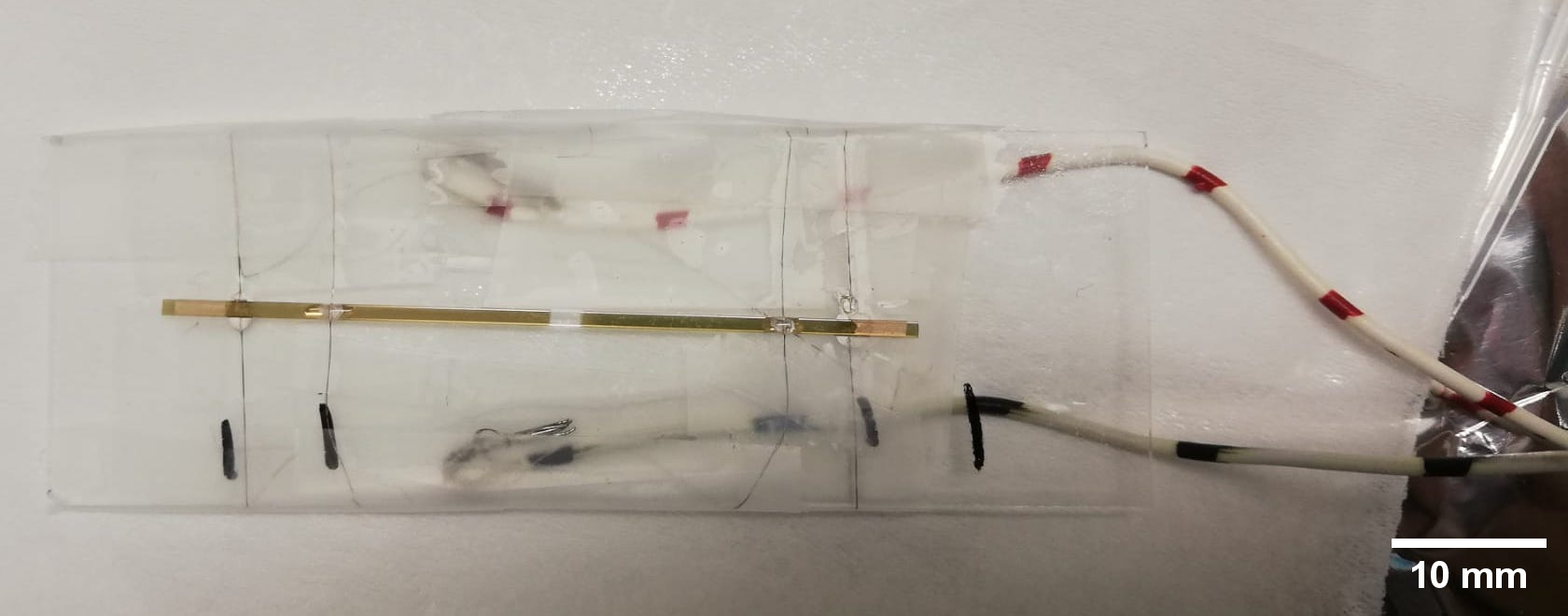}
\caption{Picture of the capillary cell of the dielectrophoretic bottle where the colloidal suspension is loaded. The golden surfaces are the electrodes}
\label{Fig:setup picture}
\end{figure}

In order to achieve a slow compression of the colloidal suspension, we construct an electric bottle with a ‘slit-like’ geometry as shown in Suppl.~\Fig{e-setup}. A picture of the electric bottle setup is depicted in Suppl.~\Fig{setup picture}. The chamber for the colloidal suspension consists of a glass capillary of cross-section 0.1$\times$1.0 mm, and 4 cm in length (Vitro, UK). The top plate of the chamber is in contact with two golden electrodes, separated by 2 mm. Such electrodes are made by sputter coating the capillary with chromium ($\sim$2 nm) and gold ($\sim$8 nm), where chromium is used for better adhesion of the golden layer. In order to achieve the desired separation distance between the electrodes, we use another capillary of measured external width of 2 mm as a mask. We repeat the same procedure on the bottom plate, to produce two additional electrodes, aligned with those on the top plate.
	
For the electrical contacts with the gold electrodes, we use silver paint (Jeol) and thin thermocouple alloy wires (diameter 50 $\mu$m, Goodfellow).
After connecting the electrodes of the top surface to each other (and similarly the electrodes of the bottom surface), we wrap the free end of the thermocouples around a standard electrical wire to further connect the electrodes to a wide band voltage amplifier. The gold-coated capillary cell is glued on top of a 1.5 mm thick standard microscopy slide (Menzel Gl\"{a}ser, Germany) with no. 71 UV-curing optical adhesive (Norland), for extra support and easy mounting on the stage of the microscope. Hence, we fill the cell with the colloidal suspension, already equilibrated in the stock vial for 8 days, and we seal both ends of the capillary with UV-curing optical adhesive (Norland no.68). The glue is cured with a handheld UV-light source ($\lambda=$350-370 nm, UVP, Switzerland). To avoid the fluorescence bleaching of particles during the curing process, the capillary is covered with an aluminum foil. The suspension is then left to further equilibrate in the electric bottle sample cell for 3 days, before turning the electric field on. For the generation of the electric field, we use a function generator (Agilent, Model 3312 OA) and a wide band voltage amplifier (Krohn-Hite, Model 7602M). We apply an AC electric field with strength $E_{rms} =$ 0.2, 0.25 or 0.3 V$/\mu$m at a frequency of 1 MHz. The field strength and the frequency are measured by an oscilloscope (Tektronix, Model TDS3052). The dielectric constant  ($\epsilon_p \approx 2.6$) of the PMMA particles is smaller than the dielectric constant of the suspending medium ($\epsilon_m \approx 3.4$). Hence, we exploit the negative dielectrophoretic forces \cite{PhysRevLett.96.015703} to control the particle concentration by slowly compressing to the slit location where the electric field is zero. By applying the electric field, we slowly increase the particle concentration starting from a low particle concentration ($\sim$ 20~wt\%).
	
\subsection{Microscopy}
\noindent We study the particle dynamics using confocal laser scanning microscopy (Leica TCS SP8 inverted) equipped with 63X (HC PL APO CS2, NA 1.40 OIL) objective lens, in the middle of the slit, in order to minimize the edge effects. We record the data sets with 512$\times$512 and 1024$\times$1024 pixels. Note that we first compress the colloidal suspension for 5 days. Next, we let the system relax for about 24 hrs before further recording the data sets for the analysis.

\subsection{Simulations}
	
The Wahnstr\"{o}m Lennard-Jones (LJ) pair-potential can be used to model a general binary metallic glass~\cite{wah91,ped10,der17,der20}, producing qualitatively similar results when compared to more accurate material-specific force models taking into account the unsaturated nature of the metallic bond~\cite{zem15}. Its parameterization is distinguished by the interaction between atom types differing in length scale, but not energy scale, which makes it also an appropriate force model for the presently used experimental colloidal system. The pairwise interaction potential between two particles $i$ and $j$ is then described as $V_{ij}(r)=4\varepsilon[(\sigma_{ij}/r)^{12}-(\sigma_{ij}/r)^{6}]$. Here the subscripts $i$ and $j$ denote the type of the two particles,  small ($a$) or large ($b$). The parameter $\sigma_{ij}2^{1/6}$ indicates the center-to-center distance below which the particles $i$ and $j$ repel each other. For comparison to the experimental systems, following the Wahnstr\"{o}m LJ parameterization, we take $\epsilon=1$, and set the ratio  $\sigma_{aa}/\sigma_{bb} = \alpha$, namely the ratio of the experimental particle diameters. In particular, we take the pairs $\sigma_{aa}=2.1~\mu$m, $\sigma_{bb} = 3~\mu$m ($\alpha=0.84$) and $\sigma_{aa}= 3.008~\mu$m, $\sigma_{bb}=3.576~\mu$m ($\alpha = 0.7$), while $\sigma_{ab}=(\sigma_{aa}+\sigma_{bb})/2$. The choice is motivated by the fact that the relevant part of the potential determining geometric frustration is the repulsive component \cite{Chandler}. The validity of such a choice is confirmed by the fact that it allows to fit the position of the first peaks of the partial radial distribution functions for small and large particles, separately considered, within a $\sim1$\% error. Mass ratios are set to the resulting spherical volume ratios. It is noted that when the particle radii of this bi-dispersed (binary) system are convoluted by a Gaussian with a width of up to 10\%, the introduced polydispersity is found not to affect the results found in the present work.

For the simulations, a periodic simulation cell of 32000 particles is used with compositions $\chi_s$ matching the experimental ones. All simulated glass structures are obtained through a linear temperature quench from a well equilibrated (converged mean and standard-deviation of energy per atom) high temperature liquid, at fixed zero-pressure. This is done by first performing a high temperature liquid equilibration at the reduced temperature $\hat{T}=k_{\mathrm{B}}T/\varepsilon=2.82$ at the fixed volume per atom $0.767\sigma_{aa}^{-3}$, followed by further equilibration at fixed zero pressure at the reduced temperature $\hat{T}=0.86$ with the average volume per atom equalling 0.865~$\sigma_{aa}^{3}$. Details of the chosen sample preparation protocol can be found in \cite{der21a,der22}. For the linear temperature quench, three quench rates are simulated at fixed zero pressure, spanning three-orders of magnitude: $\dot{T}_{n}=\dot{T}_{0}/10^{n}$, with $n = 1,2,3$ and $\dot{T}_0=1.34$. All simulations are done using the LAMMPS atomistic simulation platform with the temperature being controlled via a Noose-Hoover thermostat and (fixed zero) pressure being controlled via a Noose-Hoover barostat~\cite{tho22}. Thus the particle density in simulations is controlled by the constraint of zero pressure, and as a result the density will rise as the temperature decreases due to thermal contraction.

\subsection{Comparison between experiments and simulations}	
Suppl.~\Fig{T-energy, T-volume sim} displays the cohesive energy and volume per atom for the three quench rates for the system with diameter ratio $\alpha = 0.7$, showing that, as the temperature decreases, both energy and volume decrease. The glass transition temperature regime is characterized by a more rapid decrease in these quantities, allowing for an estimate of the glass transition temperature, as described below and in Suppl.~\Fig{NewFigSOM1}a.

To map the colloidal systems onto the thermal atomistic one, for each quench rate $\dot{T}_n$, we determine an effective temperature $T_n^*$ of the arrested colloidal system, based on the mapping quantitative matching of the experimental and simulated $(N_{4},N_{5},N_{6})$ populations. To this end, we identify the simulated temperature $T_n^*$ corresponding to a minimum in the root-mean-square (RMS) residual of the detected populations frequencies (see Suppl.~\Fig{Min residuals}).
For the two slowest quench rate the effective temperature $T_n^*$ depends little on the quench rate, because in the system is in the meta-equilibrium of the under-cooled liquid. For the fastest quench rate, there is a statistically meaningful difference. However for this quench rate, the under-cooled liquid has not achieved a meta-equilibrium at all higher temperatures. Given the meta-equilibrium of the under-cooled liquid, we mainly consider the slowest quench rates throughout the work, with a focus on the slowest simulations. In order to evaluate the proximity of the system to the glass transition, we normalize the detected effective temperatures with respect to the glass transition temperature $T_{\mathrm{g}}$.

An estimate of the glass transition temperature can be made by extrapolating the linear part of the low-temperature glass and the high-temperature under-cooled liquid, and taking the value of $T_{\mathrm{g}}$ as their intersection. Suppl.~\Fig{NewFigSOM1}a displays an example of such a construction for a system with radii ratio $\alpha = 0.7$ and composition $\chi_s = 0.65$, giving a glass transition temperature $T_{\mathrm{g}}\approx 0.33$. Suppl.~\Fig{NewFigSOM1}b shows the estimated values of the glass transition temperature for different compositions $\chi_s$.

\subsection{Simulated structural dynamics}	

To investigate the dynamical regime of the under-cooled liquid we consider the intermediate scattering function (ISF) evaluated at reciprocal space magnitudes probing the peak region of the first diffraction ring. In particular we perform isothermal simulations for a range of temperatures within the meta-equilibrium of the under-cooled liquid regime, and evaluate the instantaneous scattering amplitude at time $t$:
\begin{equation}
S_{\mathbf{q}}(t)=\frac{1}{\sqrt{N}}\sum_{i}\exp\left(i\mathbf{q}\cdot\mathbf{r}_{i}(t)\right).
\end{equation}
These values are then used to determine the temporally averaged correlation function $\mathrm{ISF}(t)=\left\langle S_{\mathbf{q}}(t_{0})S^{\dag}_{\mathbf{q}}(t_{0}+t)\right\rangle_{t_{0}}$, which is a real quantity under (meta-)equilibrium conditions. It is noted that we employ a form-factor of unity for both particles, giving equal emphasis to both particle sizes.

The viscosity for the above temperature range may obtained via the stress-stress temporal correlations function. In particular we consider the correlation between the off-diagonal components of the stress tensor, namely the shear $\sigma_{xy}$, $\sigma_{xz}$ and $\sigma_{yz}$ components, defined by the formula
\beq
\sigma^{\mu\nu}=\frac{1}{3V}\sum_{\langle ij\rangle}V'(r_{ij})\frac{r_{ij}^{\mu}r_{ij}^{\nu}}{r_{ij}},
\eeq
where $V$ is the volume of the periodic system, $r_{ij}^{\mu}$ is the $\mu$th component of the distance between particles $i$ and $j$, and $\langle ij\rangle$ is a summation over all interacting pairs of particles.
For a meta-equilibrium system, an estimate for the shear viscosity can be obtained via Green-Kubo linear-response theory \cite{gre54,kub57} as
\begin{equation}
\eta^{\mu\nu}=\frac{V}{k_{\mathrm{B}}T}\int_{0}^{\infty}\mathrm{d}t\left\langle\sigma^{\mu\nu}(t_{0})\sigma^{\mu\nu}(t_{0}+t)\right\rangle_{t_{0}}. \label{eqnEta}
\end{equation}
The determined shear viscosity is derived from an average of $\eta^{xy}$, $\eta^{xz}$ and $\eta^{yz}$.

For both the intermediate scattering function and viscosity, sufficiently long simulations were performed to obtain converged values for the under-cooled liquid regime considered. For each of the isotherms, crystallization was not observed and thus Eqn.~\ref{eqnEta} (with the upper limit being replaced by a time larger than the timescale of the stress fluctuations) remains valid, probing the restricted ergodicity of the under-cool liquid. At lower temperatures, approaching the measured $T_{\mathrm{g}}$ the under-cooled liquid exits its meta-equilibrium state and relaxation occurs at the time-scale of the simulations. For this latter temperature regime, both the $\mathrm{ISF}(t)$ and $\eta^{\mu\nu}$ cannot be reliably calculated for the timescales simulated.

\subsection{Image analysis}

As a consequence of the secondary excitation of the NBD by the emission of the Nile Red (or RITC) fluorescent dye, the experiments provide two types of microscopy images for the same colloidal configuration:
one with only the small particles and one with both the small and the large particles. We first subtract the image with the small particles from the corresponding image with large and small particles, thus
obtaining an image with the large particles only. We then separately detect the position of small and large particles with TrackMate \cite{tin17}. To verify that the detection has subpixel precision, we check the uniform nature of the probability density function (PDF) of the residuals $x^{pxl}_i - \text{floor}(x^{pxl}_i)$, $y^{pxl}_i - \text{floor}(y^{pxl}_i)$, $z^{pxl}_i - \text{floor}(z^{pxl}_i)$, where $x^{pxl}$, $y^{pxl}$, $z^{pxl}$ are the particle coordinates expressed in pixels, and $\text{floor}(x)$ denotes the nearest integer less than $x$. We additionally note that the analysis of SU(2) is quite insensitive to local deformations, while it is sensitive to full misdetections, varying the bond order of a particle. Optical inspection on randomly selected volumes containing $\sim$500 particles reveals a misdetection occurrence rate lower than $1\%$. \\
	
\subsection{Structural analysis}

To determine the local SU(2) structural motifs, i.e. the local bonding classes \cite{nel83a,nel83b}, a modified radical Voronoi tessellation is first used to identify nearest neighbour bonds. The modification removes bonds generated from the standard radical Voronoi algorithm to produce a tessellation of less distorted tetrahedra with resulting bond-orders similar to those obtained using a common neighbour analysis~\cite{der20a}. The analysis is performed with the atomistic visualization software Ovito~\cite{stu10} via an Ovito modifier~\cite{GITHUB}.The bond-order population $(N_{4},N_{5},N_{6})$ of each atom, the SU(2) bonding class, is then determined by enumerating these nearest neighbour bonds and their bond-order, i.e. the number of common neighbours or equivalently the number of edges of the shared Voronoi face of the corresponding atom pair. The work of Ref.~\cite{der20a} also established that the orientational arrangement of the bonds generally follows the geometrical structure predicted by Refs. \cite{nel83a,nel83b}.

The simulated SU(2) motif populations shown in Fig.~1 encompass also the ``other'' category. These consist of local structural motifs not identified as those listed, and generally represent higher local energy and volume, and hence more liquid-like regions. Due to the low statistics of each of the local motifs included in the "other" category, extending the histogram to such motifs reveals little statistically relevant information for the sample sizes considered. Instead, Suppl.~\Fig{Bond order-coordination} shows the normalized histograms of both all encountered bond-orders and coordination numbers for experiments and simulations. Deviations between them in bond order mainly occur outside the range of order-4, -5 and -6, which also correspond to extremal values of coordination. It is these values that constitute the bulk of the ``other'' category, often representing open-volume regions (low coordination $Z<8$) and high density regions (highly frustrated, with high coordination $Z>20$) that are mainly seen in experiment.
		
A recent approach to investigate local and medium range order is to directly calculate the 4-particle density correlation function \cite{yua21,sin23} by using three mutually nearest neighbour particles to define a coordinate system with which to measure the location and orientation of any fourth particle. Here the first particle defines the origin, and the remaining two particles define a Cartesian plane. Upon decomposing the anisotropic radial histogram of the particle density $\rho(r,\theta,\varphi)=\sum_{l=0}^{\infty}\sum_{m=-l}^{l}\rho_{l}^{m}(r)Y_{l}^{m}(\theta,\varphi)$ with respect to spherical harmonics as a function of radius, a medium range order can be seen in the $l=6$ harmonic, $S_{\rho}(l,r)=\left[(2l+1)^{-1}\sum_{m=-l}^{l}|\rho_{l}^{m}(r)|^{2}\right]^{1/2}$ which decays relatively slowly with distance.


\clearpage

\begin{figure*}
\includegraphics[width=0.5\linewidth]{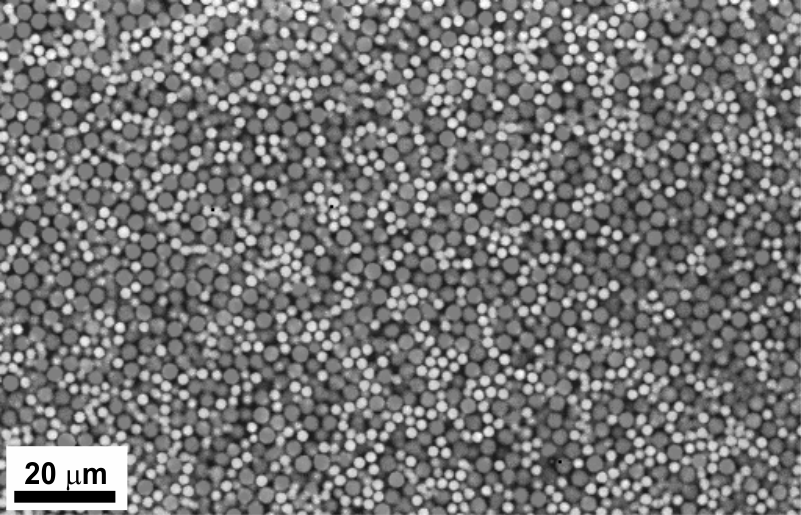}
\caption{Arrested binary colloidal system, formed by PMMA particles with diameter 3~$\mu$m and 2.1~$\mu$m, dielectrophoretically compressed with $E_{rms}$= 0.25 V/$\mu$m up to a final volume fraction $\Phi\sim0.52$.}
\label{Fig:Arrested experiment}
\end{figure*}

\begin{figure*}
\includegraphics[width=0.5\linewidth]{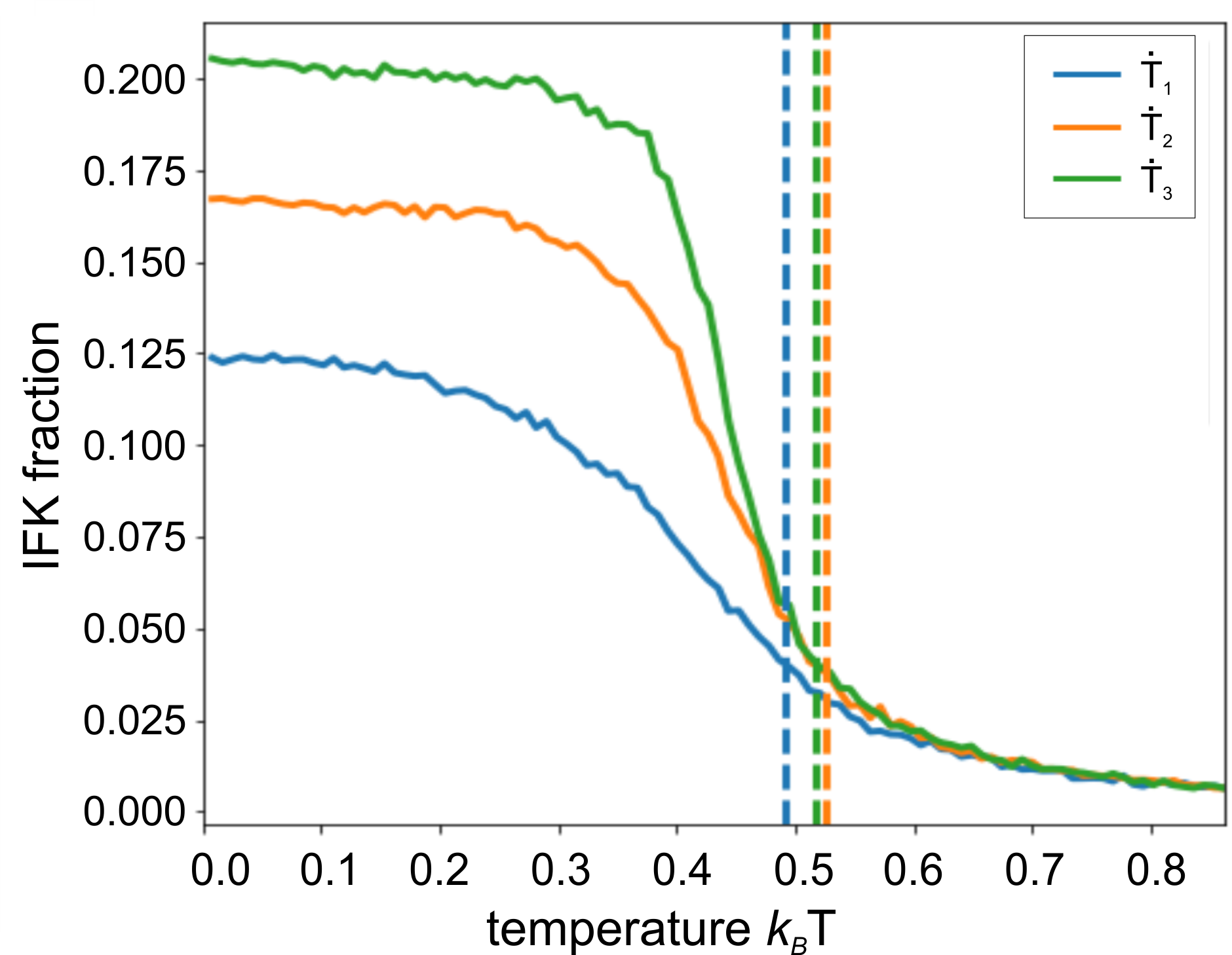}
\caption{Plot of the fraction of atoms with a local icosahedral or Frank-Kasper (IFK) bonding topology versus quench temperature, for atomistic simulations of a LJ binary system, for quench rates  $\dot{T}_n=\dot{T}_0/10^n$ with $n$=1,2,3. The vertical lines indicate the temperatures corresponding to the best fit between instantaneous structures in the simulated atomistic system and the experimental colloidal system, namely the minima in Suppl.~\Fig{Min residuals}. The radii ratio is $\alpha = 0.7$.}
\label{Fig:T-IFK sim}
\end{figure*}

\begin{figure*}
\centering
\includegraphics[width=1\linewidth]{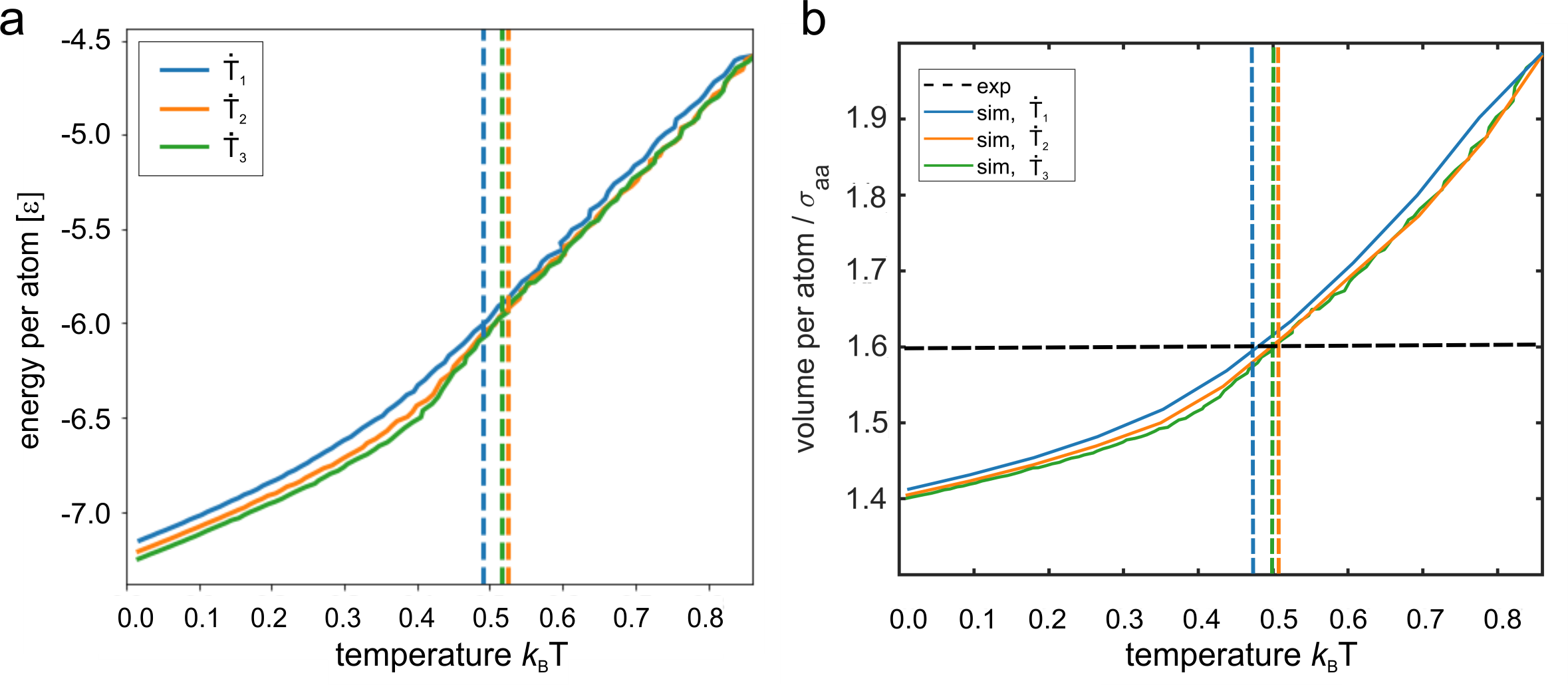}
\caption{Plot of a) cohesive energy and b) volume per atom as a function of temperature in atomistic simulations of a temperature-quenched LJ binary system. The colours indicate three different quench rates $\dot{T}_{n}=\dot{T}_{0}/10^{n}$, with $n = 1,2,3$. The horizontal dashed black line corresponds to
the arrested colloidal experimental system. The vertical lines indicate the temperatures corresponding to the best fit between instantaneous structures in the simulated atomistic and experimental colloidal system, namely the minima in Suppl.~\Fig{Min residuals}. The radii ratio is $\alpha = 0.7$.}
\label{Fig:T-energy, T-volume sim}
\end{figure*}

\begin{figure*}
\centering
\includegraphics[width=0.5\linewidth]{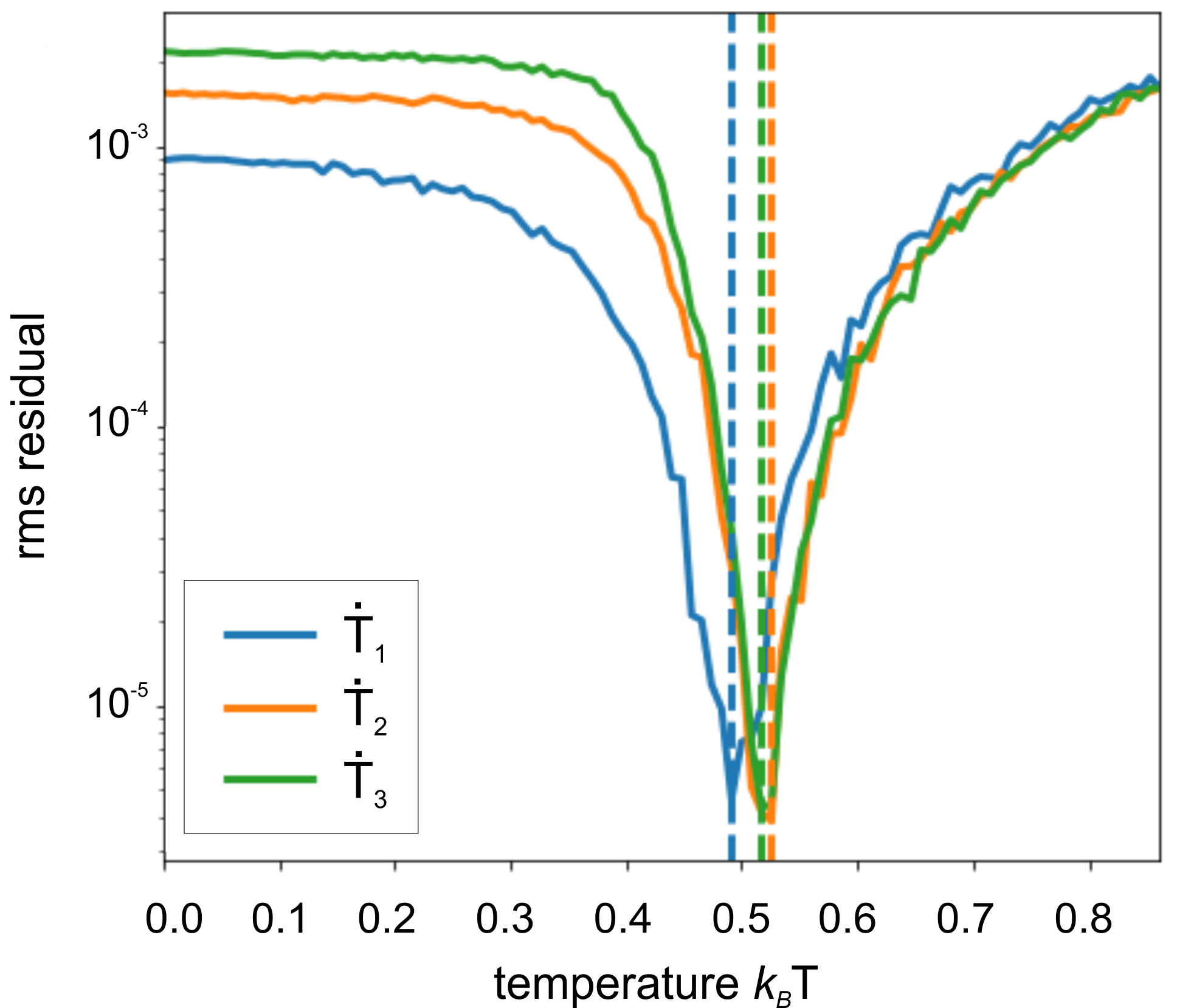}
\caption{Plot of the sum of residuals in the relative populations of SU(2) local motifs for a simulated thermally quenched LJ atomistic system with respect to an experimental colloidal athermally compressed system in the arrested state. The colours indicate three different quench rates $\dot{T}_{n}=\dot{T}_{0}/10^{n}$, with $n = 1,2,3$. The vertical dashed lines represent the effective temperatures $T_n^*$ for which the residual are minimal. The radii ratio is $\alpha = 0.7$.}
\label{Fig:Min residuals}
\end{figure*}

\begin{figure*}
\includegraphics[width=0.9\linewidth]{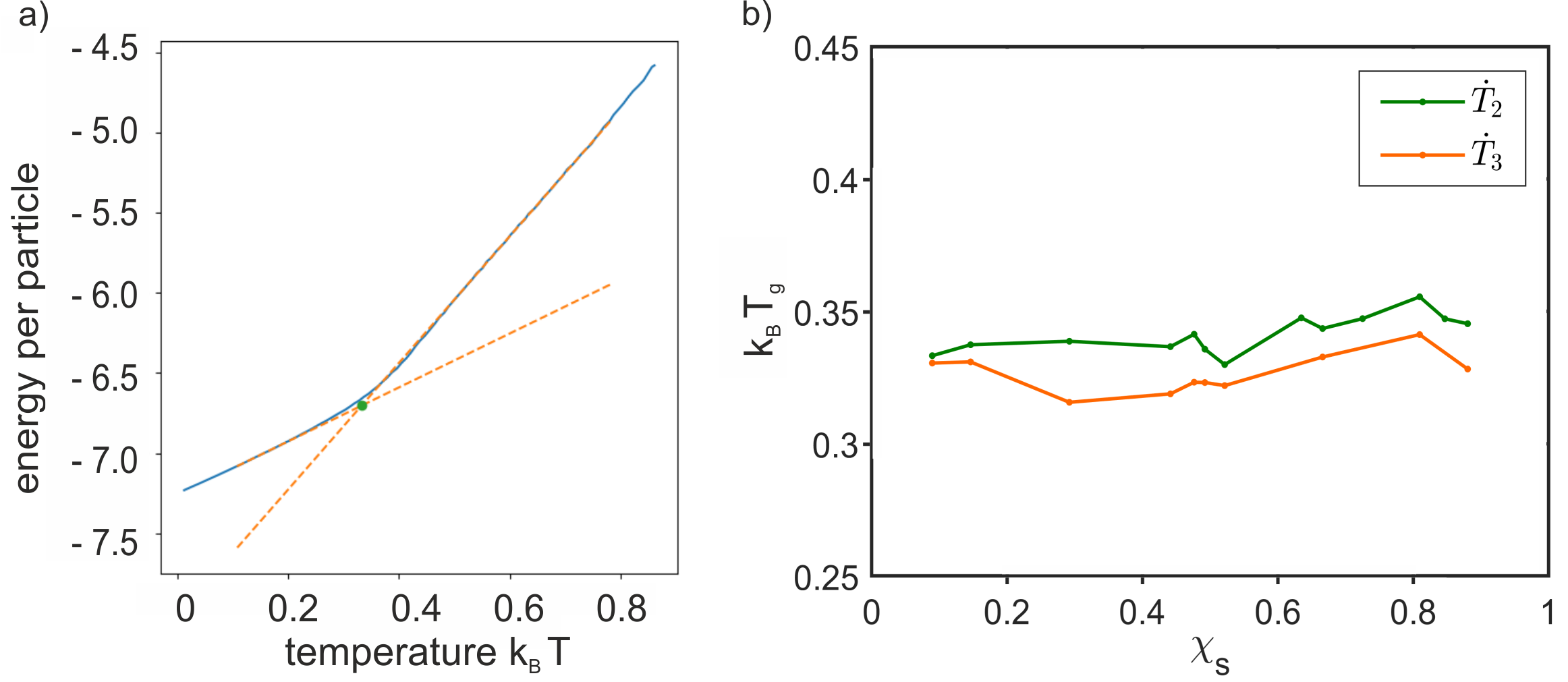}
\caption{a) Example of low and high temperature extrapolation method of energy versus temperature to estimate the glass transition temperature. b) Glass transition temperature estimate based on simulated quench rates $\dot{T}_{n}=\dot{T}_{0}/10^{n}$, with $n = 2,3$ as a function of molar composition for a binary system with a radii ratio $\alpha = 0.84$. }
\label{Fig:NewFigSOM1}
\end{figure*}

\begin{figure*}
\includegraphics[width=0.7\linewidth]{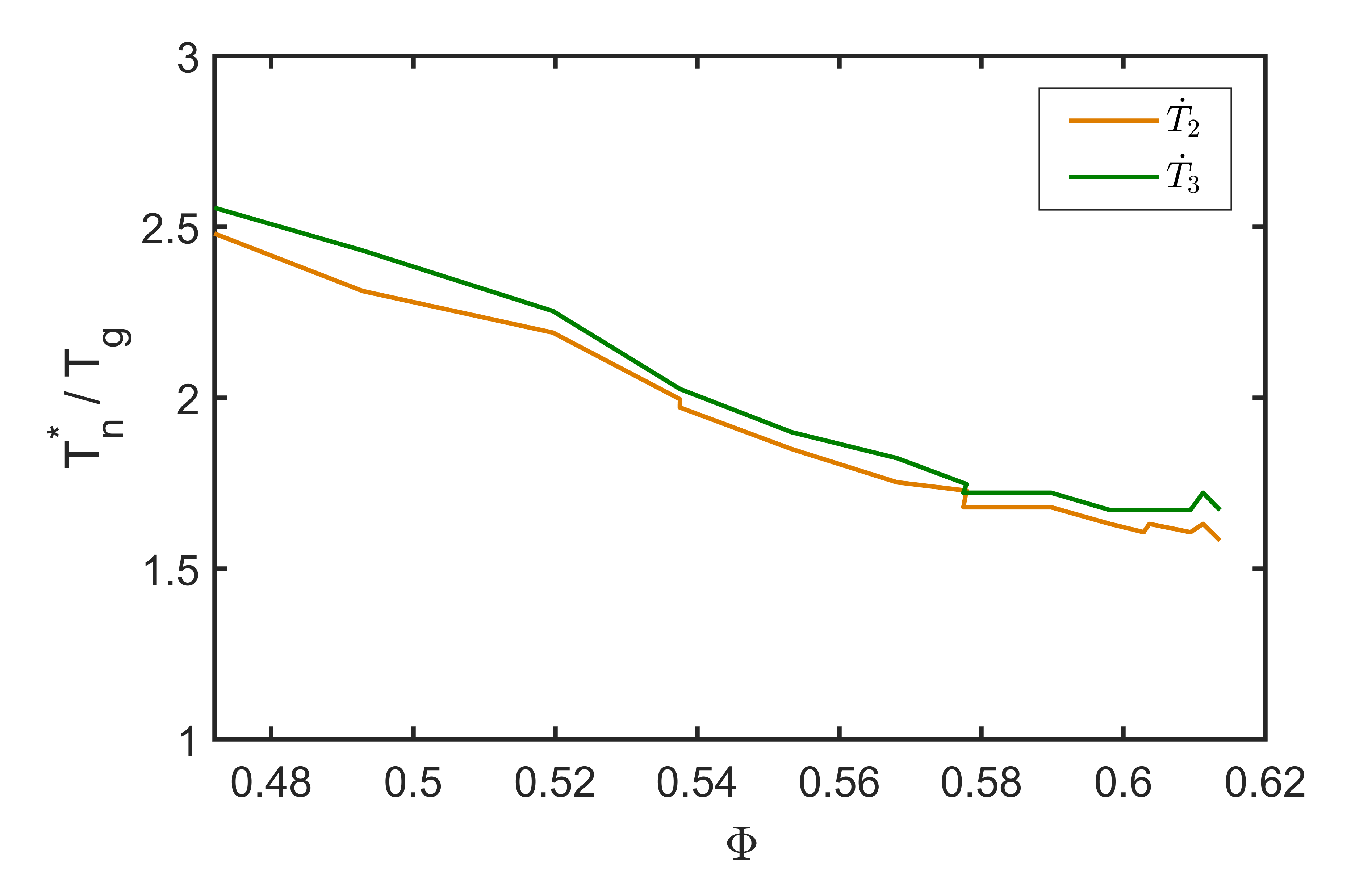}
\caption{Equivalent temperature $T_n^*$ versus volume fraction $\Phi$, for a compressed binary colloidal system, calculated with respect to two simulated quench rates $\dot{T}_n$, with $n=2,3$. The system has a radii ratio $\alpha = 0.84$, composition $\chi_s = 0.85$ and the compression is achieved with $E_{rms} = 0.2$ V/$\mu$m.}
\label{Fig:Tmap different quench}
\end{figure*}

\begin{figure*}
\includegraphics[width=0.62\linewidth]{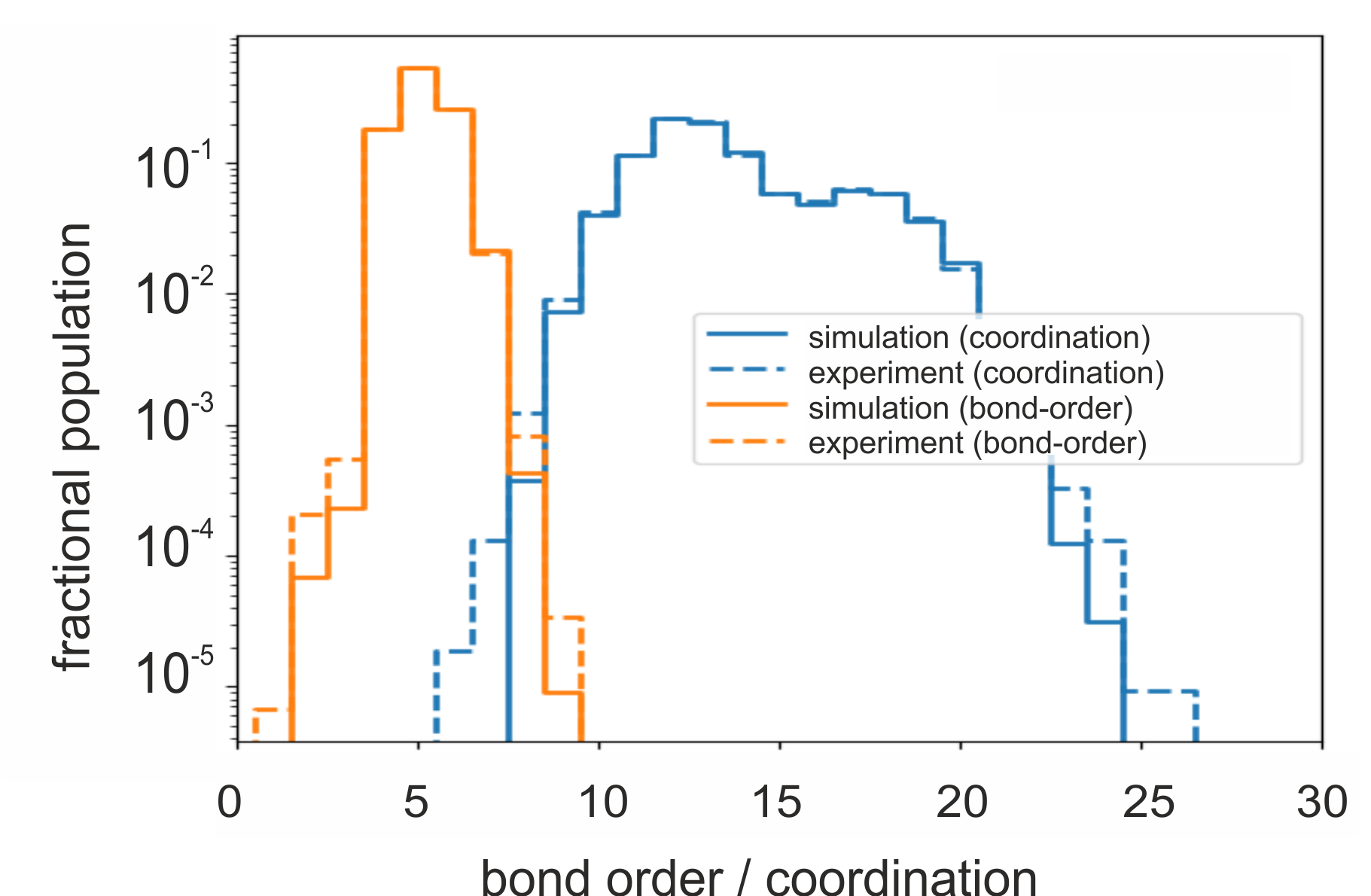}
\caption{Plot of bond order and coordination normalized histograms for the colloidal system of Fig.~1 and the corresponding simulated thermal configuration.}
\label{Fig:Bond order-coordination}
\end{figure*}

\begin{figure*}
\centering
\includegraphics[width=0.6\linewidth]{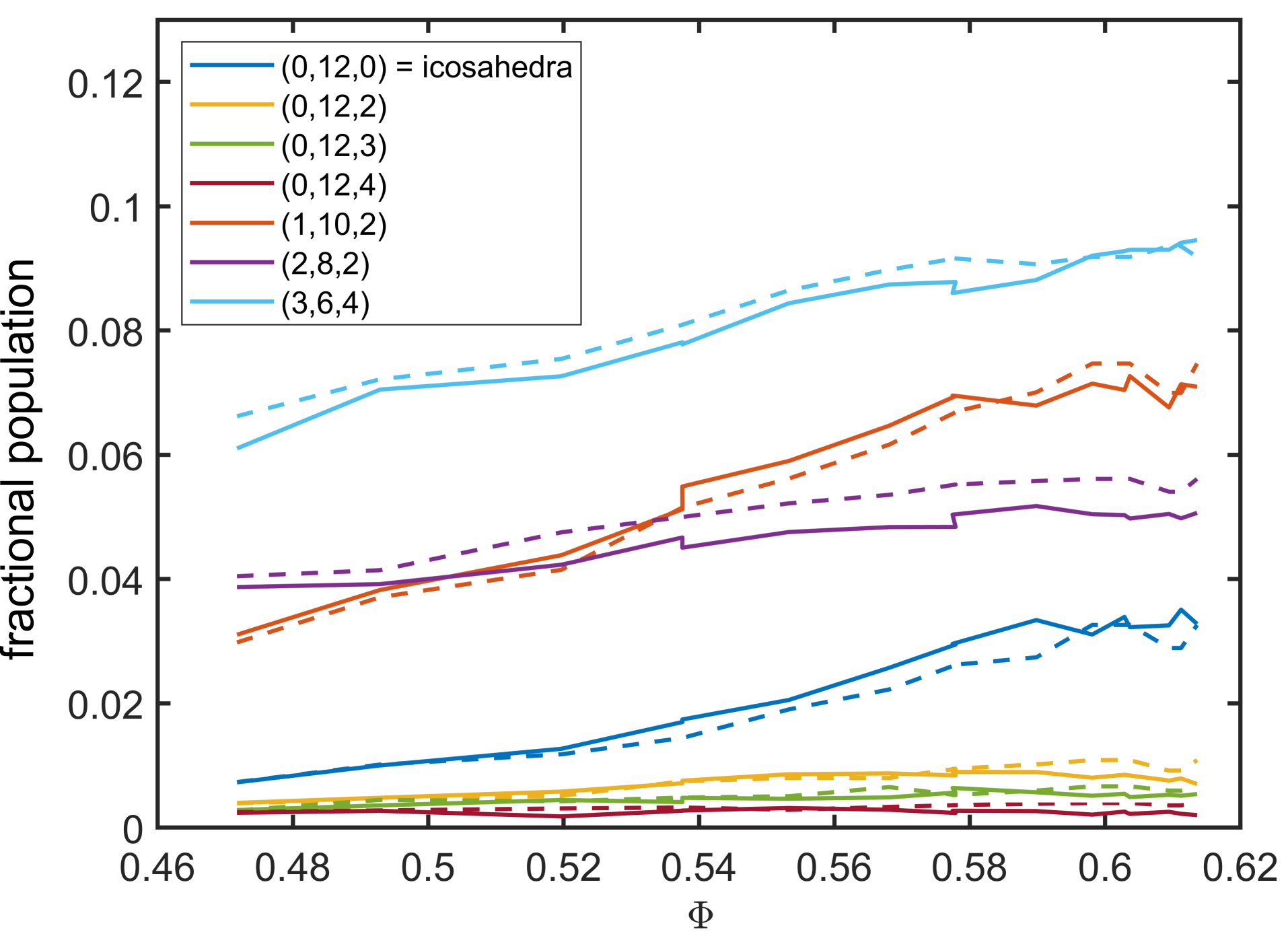}
\caption{Evolution of the populations of the main SU(2) local motifs, for a colloidal system compressed with $E_{rms} = 0.2$ V/$\mu$m (solid lines) and the corresponding simulated LJ atomistic system achieved with thermal quench rate $\dot{T}_{3}=\dot{T}_{0}/10^{3}$ (dashed lines). The radii ratio is $\alpha=0.84$ and the volume fraction of small particles is $\chi_s=0.85$. }
\label{Fig:Main classes exp-sim}
\end{figure*}

\begin{figure*}
    \centering
	\includegraphics[width=0.6\linewidth]{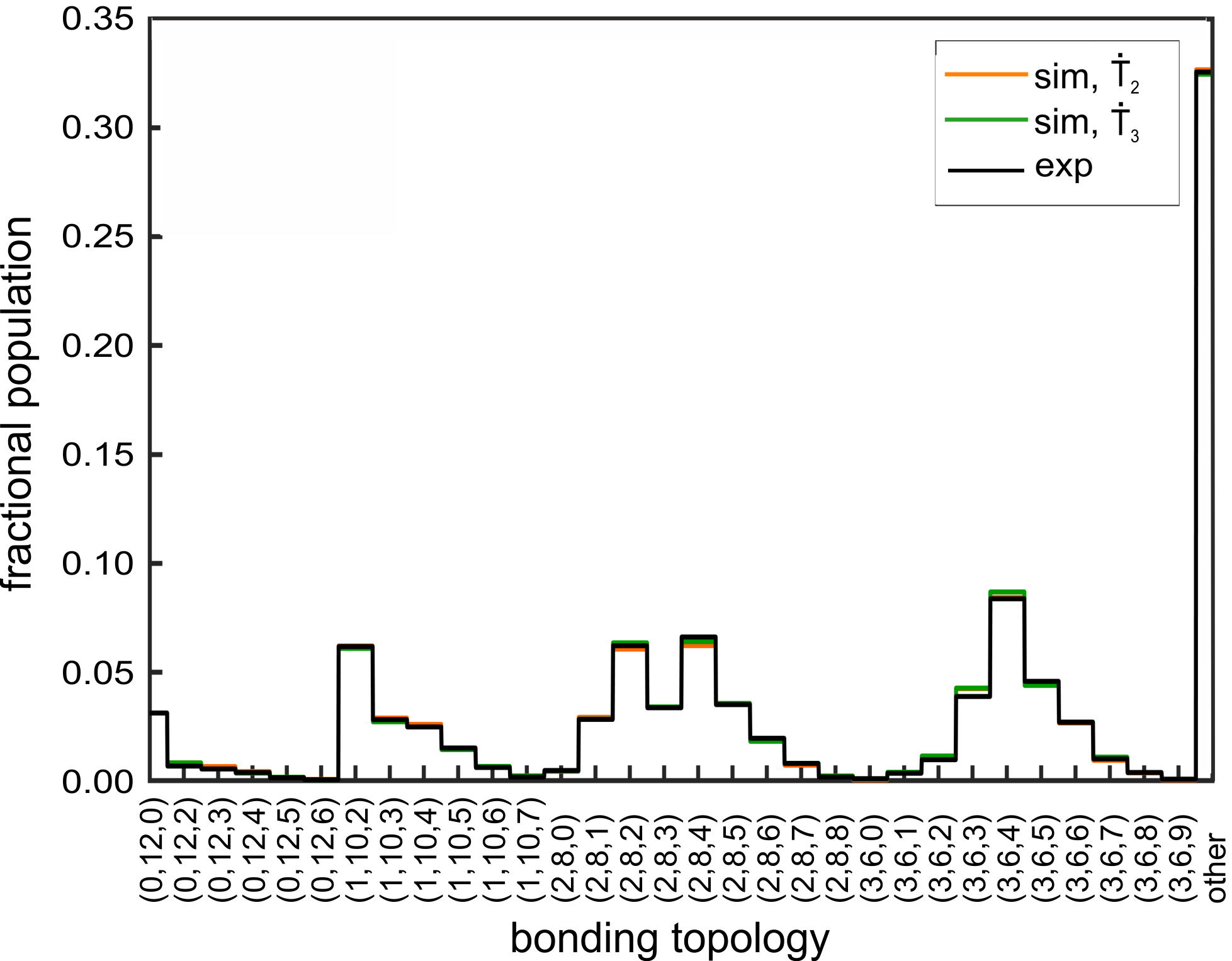}
	\caption{Comparison of the Nelson local motifs populations normalized by the total particle number, for an arrested colloidal experiment and atomistic simulations of a thermally quenched LJ system at the effective temperatures $T_n^*$, for different quench rates $\dot{T}_{n}=\dot{T}_{0}/10^n$, with $n = 2,3$. The system has $\alpha = 0.84$, $\chi_s=0.65$ and it is experimentally compressed with $E_{rms} = 0.3$ V/$\mu$m.}
	\label{Fig:Nelson classes, sim-exp 2}
\end{figure*}

\begin{figure*}
\centering
\includegraphics[width=0.65\linewidth]{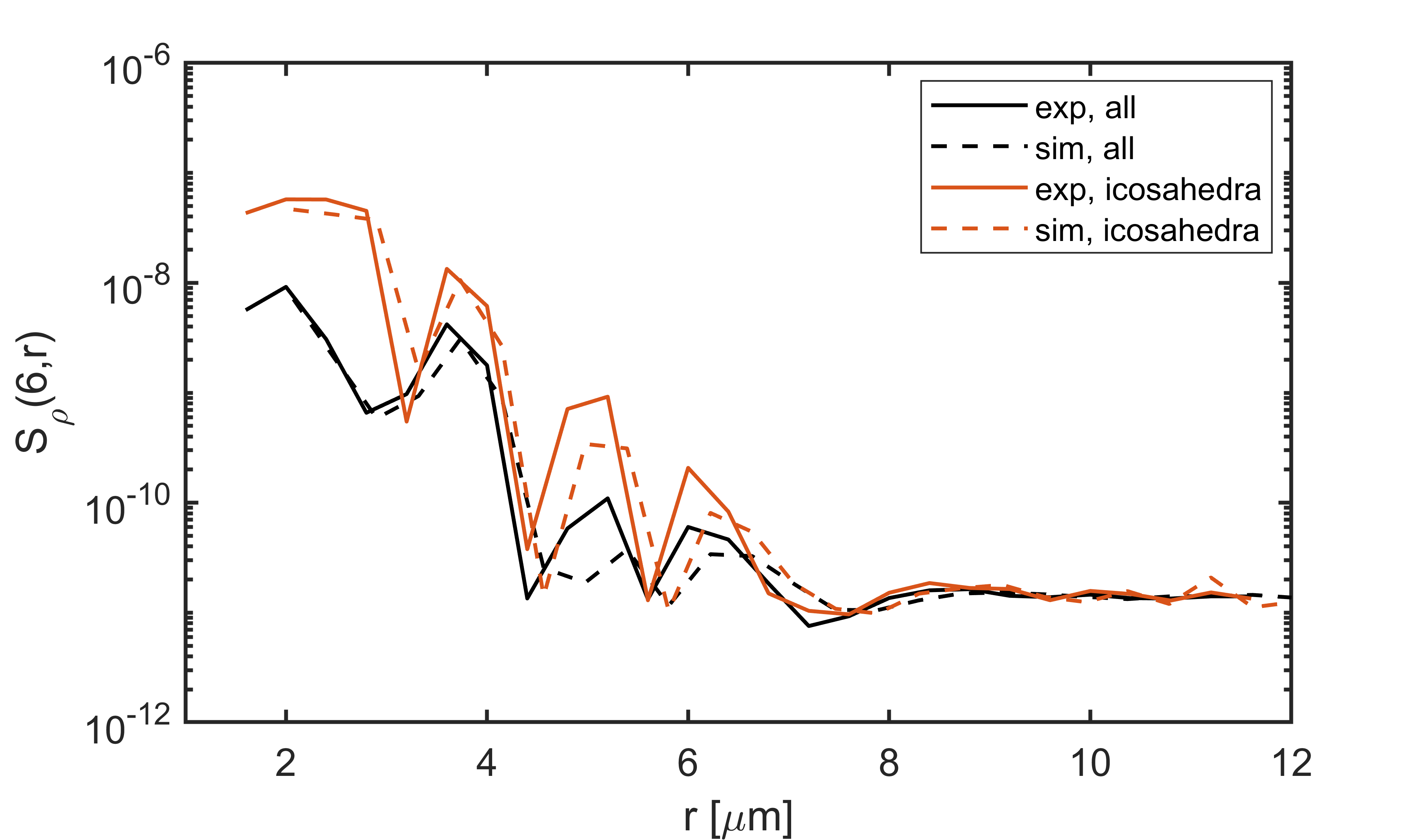}
\caption{Square root of the angular power spectrum $l=6$ versus radial distance, for all particles (black) and for icosahedral-centered particles (red), in experiments (solid lines) and simulations (dashed lines), for the system of Suppl.~\Fig{Nelson classes, sim-exp 2}.}
\label{Fig:Q6 2}
\end{figure*}

\begin{figure*}
\centering
\includegraphics[width=0.7\linewidth]{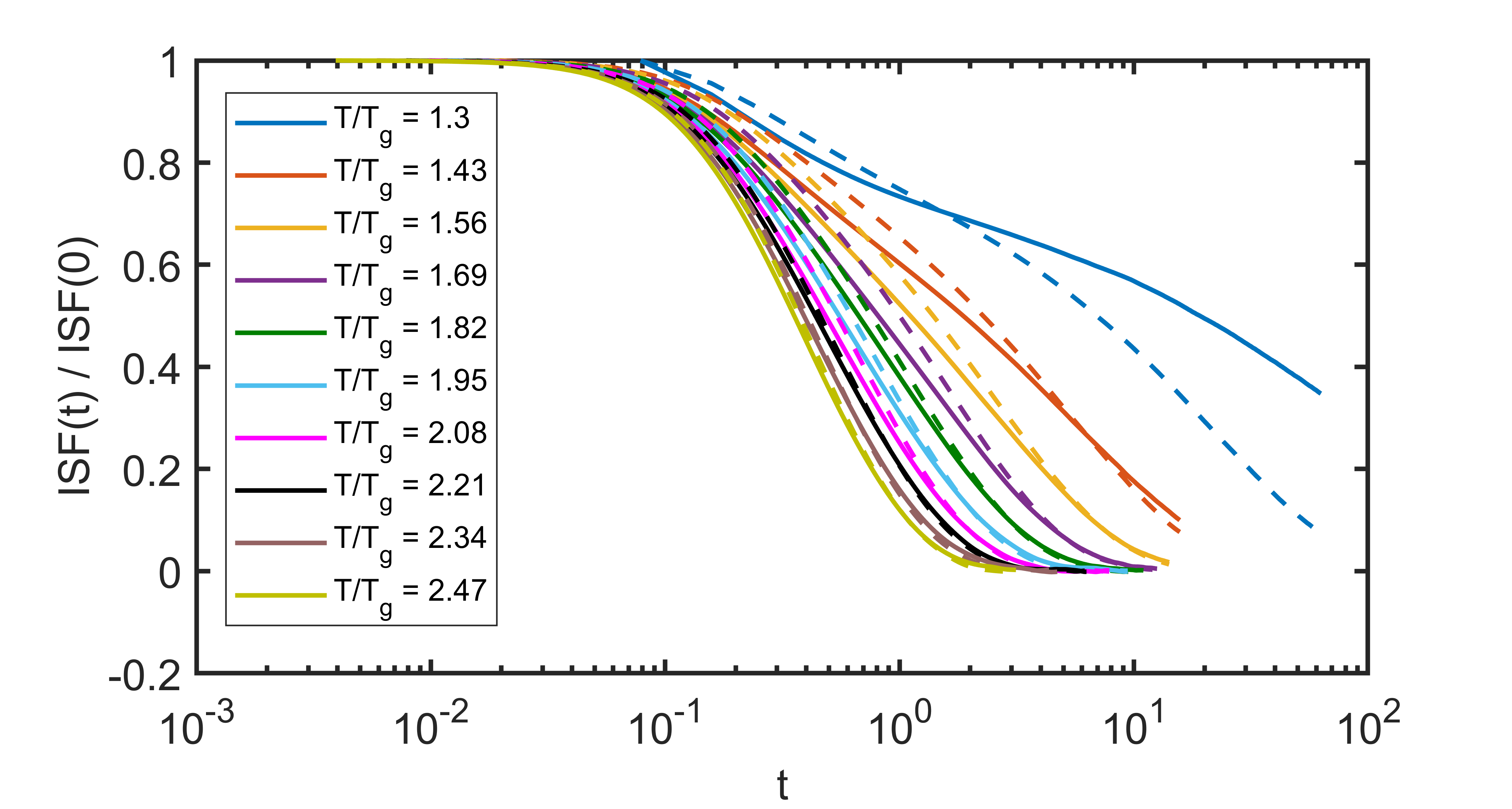}
\caption{Simulated intermediate scattering function ISF, as a function of time, for systems with radii ratios $\alpha=0.7$ (solid line) and $\alpha = 0.84$ (dashed line), at the same composition $\chi_s=0.65$. The effective temperatures of the colloidal systems are $T_3^*/T_g=1.57$ and $1.69$, respectively. Time is expressed in natural units.}
\label{Fig:ISF}
\end{figure*}

\begin{figure*}
\centering
\includegraphics[width=0.63\linewidth]{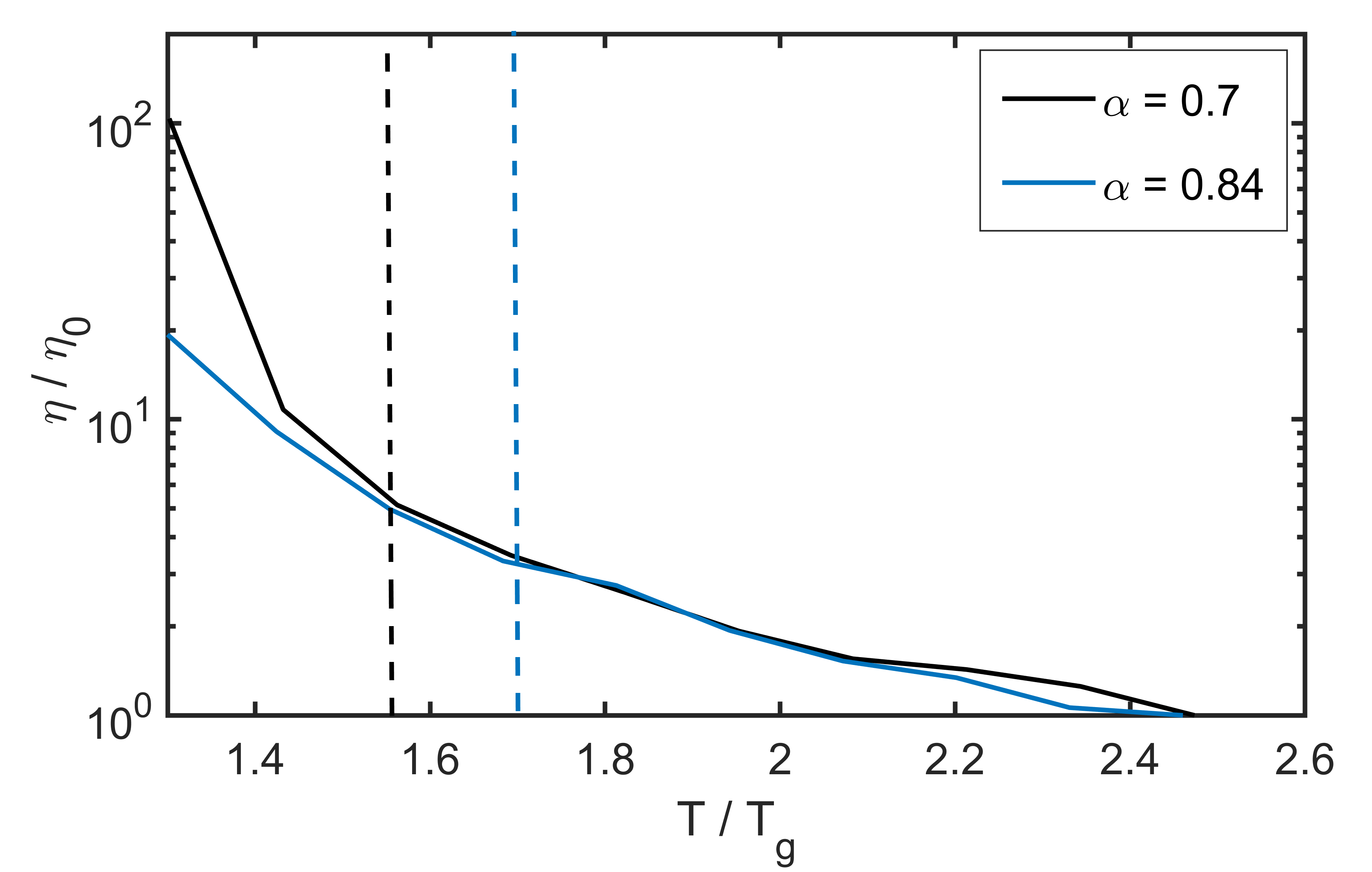}
\caption{Simulated viscosity $\eta/\eta_0$ where $\eta_0$ is the viscosity at the highest simulated temperature, for systems with $\alpha = 0.7$ (black) and $\alpha = 0.84$ (blue), at composition $\chi_s = 0.65$. The dashed vertical lines indicate the mapping temperatures $T_3^*$ in the experimentally arrested state achieved by compression with $E_{rms} = 0.25$ V/$\mu$m (black) and $0.3$ V/$\mu$m (blue).}
\label{Fig:Viscosity}
\end{figure*}

\end{document}